\documentclass[%
reprint,amsmath,amssymb,
aps,
prd,
noeprint,
nolongbibliography,
nofootinbib
]{revtex4-2}

\usepackage{graphicx}
\usepackage{dcolumn}
\usepackage{bm}

\usepackage{graphicx}
\usepackage{dcolumn}
\usepackage{bm}
\usepackage[T1]{fontenc}
\usepackage{subfigure}
\usepackage{indentfirst}
\usepackage{soul}
\usepackage[normalem]{ulem}
\usepackage{xcolor}
\usepackage{multirow}

\newcommand{\dd}{\mathrm{d}}
\newcommand{\Mpl}{M_\mathrm{Pl}}

\DeclareMathOperator{\erfc}{erfc}
\DeclareMathOperator{\arcsinh}{arcsinh}

\usepackage{hyperref}

\begin{document}

\preprint{APS/123-QED}

\title{Primordial black holes through preheating
instabilities in \texorpdfstring{$\bm\alpha$}{TEXT}-attractor models}

\author{Daniel del-Corral$^{1,2,3}$}
\email{daniel.corral.martinez@uj.edu.pl}

\author{Paolo Gondolo$^{4,5}$}
\email{paolo.gondolo@utah.edu}

\author{K. Sravan Kumar$^{6}$}
\email{sravan.kumar@port.ac.uk}

\author{João Marto$^{1,2}$}
\email{jmarto@ubi.pt}

\affiliation{$^{1}$ Departamento de Física, Universidade da Beira Interior, Rua Marquês D'Ávila e Bolama 6200-001 Covilhã, Portugal}
\affiliation{$^{2}$Centro de Matemática e Aplicações da Universidade da Beira Interior, Rua Marquês D'Ávila e Bolama 6200-001 Covilhã, Portugal}
\affiliation{$^{3}$Faculty of Physics, Astronomy and Applied Computer Science, Jagiellonian University, 30-348 Krakow, Poland}
\affiliation{$^{4}$Department of Physics and Astronomy, University of Utah, Salt Lake City, UT 84112, USA}
\affiliation{$^{5}$Department of Physics, Institute of Science Tokyo, 2-12-1 Ookayama, Meguro-ku, Tokyo 152-8551, Japan}
\affiliation{$^{6}$Institute of Cosmology and Gravitation, University of Portsmouth, Dennis Sciama Building, Burnaby Road, Portsmouth, PO1 3FX, United Kingdom}


\begin{abstract}
In this work, we explore the production of primordial black holes (PBHs) within the context of $\alpha$-attractor inflationary models, focusing on the preheating phase following inflation. During this phase, self-resonance instabilities arise due to deviations of the inflationary potential from a quadratic form. PBH formation is analyzed using three criteria: (1) the perturbation must lie within the instability band, (2) its characteristic length must exceed the Jeans length, and (3) it must have sufficient time to collapse based on the estimations of massive scalar field spherical collapse in Einstein-de Sitter universe. Based on these criteria, we calculate the PBH mass fraction using the {altered} Press-Schechter (PS) {that is generally implemented in preheating scenarios} and Khlopov-Polnarev (KP) formalism {that considers non-spherical effects}. Our results show that the {altered} PS formalism tends to overestimate PBH abundance during preheating {in} contrast to the KP formalism. We provide a detailed comparison {between these two frameworks} with observational constraints from evaporating PBHs. Notably, the {altered} PS formalism is excluded by these constraints, which are based on Hawking radiation, while the KP formalism remains viable. These findings underscore the importance of accounting for nonspherical effects and accurate collapse dynamics in studies of PBH formation during preheating.
\end{abstract}

\maketitle



\section{Introduction}\label{sec:introduction}

Primordial Black Holes (PBHs), first proposed by Zeldovich and Novikov \cite{Zeldovich:1967} and independently by Hawking and Carr \cite{Hawking:1971,Carr:1974,Carr:1975}, remain a current topic of discussion. These black holes are hypothesized to have formed from the collapse of scalar perturbations in the primordial universe, before the epoch of big bang nucleosynthesis. This early formation implies that PBHs span a wide range of masses and thus may explain a plethora of phenomena such as dark matter, the origin of structure in the universe, or the supermassive black holes at the centers of galaxies. For comprehensive reviews on PBHs, see \cite{Villanueva-Domingo:2021,Carr:2020,Carr:2021bzv}. Hawking \cite{Hawking:1974rv} theorized in 1974 that black holes radiate via a quantum process known as Hawking evaporation, with a mass-dependent timescale $t_{\text{eva}} \sim M_{\text{PBH}}^3$. For PBHs with masses below the critical value $M_{\text{crit}} \sim 5 \times 10^{14}\,\mathrm{g}$, the evaporation timescale is shorter than the age of the universe, meaning these PBHs would have evaporated by now and cannot be directly detected. However, one can constrain their abundance by using Hawking evaporation as an indirect mechanism. This is further explained in Sec.~\ref{sec:constraints}.

Regarding PBH formation, the collapse of perturbations is typically considered to occur in a radiation-dominated universe, where radiation pressure plays an essential role in halting the collapse. This allows the definition of a threshold value above which perturbations collapse. However, during epochs of reduced pressure, such as a matter-dominated phase, PBH formation could be significantly enhanced. This study explores PBH formation during the preheating phase following inflation, an era during which the universe behaves approximately as matter-dominated. This raises new challenges in modeling PBH collapse, as the absence of pressure could, in principle, enable any overdense region to collapse. We employ three criteria to restrict PBH formation, which can be found in detail in \cite{del-Corral:2023} and are summarized in Sec.~\ref{sec:pbh-formation}. It is important to note that the constraints of Fig.~\ref{fig:constraints} assume a standard big bang cosmology. The inclusion of an early matter-dominated era, such as preheating, modifies these constraints \cite{Lemoine:2000sq,PBHBeta_documentation,PBHBeta_repository}. However, this modification is negligible for the range of masses and preheating durations we consider, and mainly affects the remnants' constraints. Therefore, it is not taken into account.

In a previous work \cite{del-Corral:2023}, we investigated PBH formation during preheating in Starobinsky inflation, based on earlier studies \cite{Martin:2019nuw,Jedamzik:2010dq,Martin:2019nuw}. In this study, we extend that analysis to a broader class of inflationary models arising within the framework of supergravity theories, known as $\alpha$-attractors \cite{Kallosh:2013daa,Kallosh:2013hoa,Kaiser:2013sna,Kallosh:2015lwa}. Starobinsky inflation is a specific case of this class, corresponding to $\alpha = 1$. In $\alpha$-attractors, the parameter $\alpha$ is the key characteristic of the model, representing the curvature of the Kähler geometry in the supergravity theory \cite{Kallosh:2013yoa,Kallosh:2015,Iacconi:2023mnw,Carrasco:2015uma}. Two of the families of $\alpha$-attractors considered in this work are the so-called T- and E-models. Their inflationary potentials are given, respectively, by
\begin{subequations}\label{eq:potentials}
\begin{gather}
\label{eq:T-model}
    V_T(\phi) = 3\alpha M^2 \Mpl^2 \tanh^2\left[\frac{1}{\sqrt{6\alpha}} \frac{\phi}{\Mpl} \right],
\\\label{eq:E-model}
    V_E(\phi) = \frac{3\alpha M^2 \Mpl^2}{4} \left(1 - e^{-\sqrt{\frac{2}{3\alpha}} \frac{\phi}{\Mpl}} \right)^2.
\end{gather}
\end{subequations}
Here, $\phi$ is the scalar field, $M$ is the scalaron mass (normalized with CMB constraints), and $\Mpl$ is the reduced Planck mass\footnote{{Throughout this work we use the reduced Planck mass $\Mpl=\frac{1}{\sqrt{8\pi G}}$ in units of $c=\hbar=1$ and follow the metric signature $(-+++)$.}}. {There are multiple investigations in the literature that have derived lower bounds on the parameter $\alpha$. Ref. \cite{Iacconi:2023mnw} found constraints on $\alpha$ from the study of reheating equation of state as $\log_{10}(\alpha) = -4.2_{-8.6}^{+5.4}$ (95\% C.L.). According to \cite{Alam:2023kia}, considerations of the overproduction of a light moduli field also implied some constraints on $\alpha$ in the context of the T-model. In \cite{Krajewski:2018moi} $\alpha \gtrsim 10^{-3}$ is obtained based on the effects of heavy fields in the $\alpha$-attractor supergravity framework, and the subsequent geometric destabilization phenomenon during inflation.  In our recent study \cite{del-Corral:2025fzz}, we obtained stringent bounds 
$\log_{10}(\alpha) > -3.54$ for the T-model and $\log_{10}(\alpha) > -3.17$ for the E-model using the possible production of scalar-induced gravitational waves during preheating and applying the bound from the BBN constraint. In this work, we consider these values of $\alpha$ to study PBH production during preheating in both E- and T-models.}

\begin{figure}[htbp]
    \centering
    \subfigure[T-model]{%
    \includegraphics[width=0.75\linewidth]{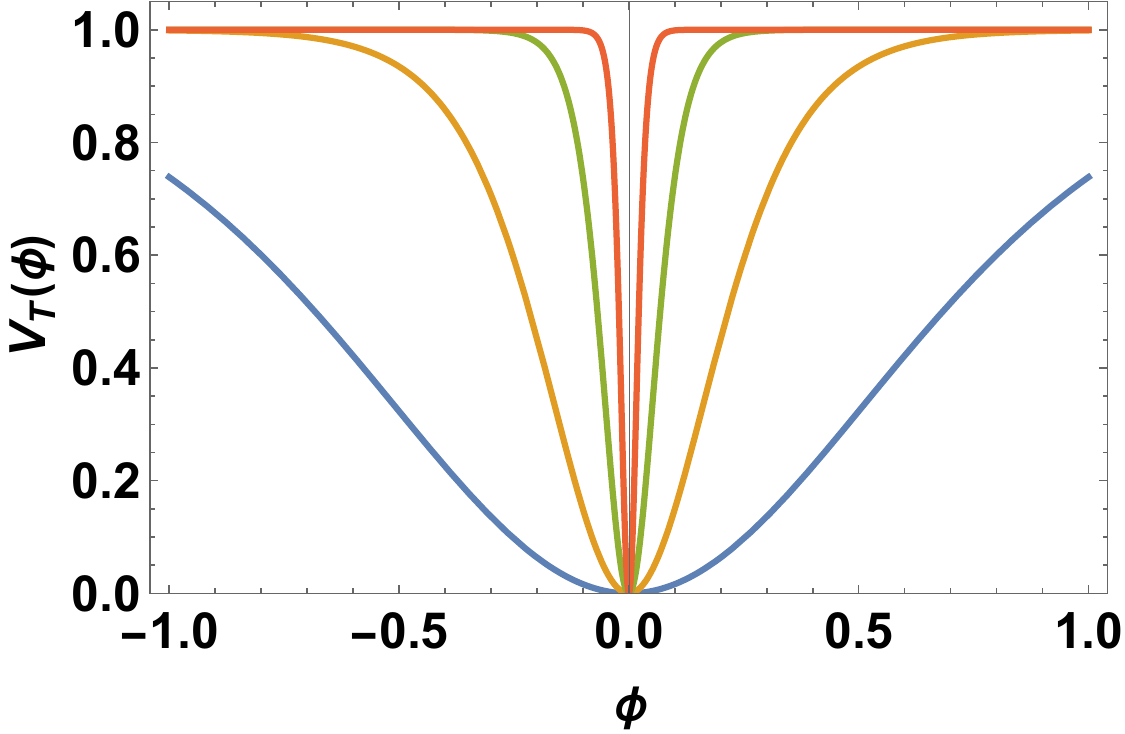}
        \label{fig:potentialsT}
    }
    \hfill
    \subfigure[E-model]{%
        \includegraphics[width=0.75\linewidth]{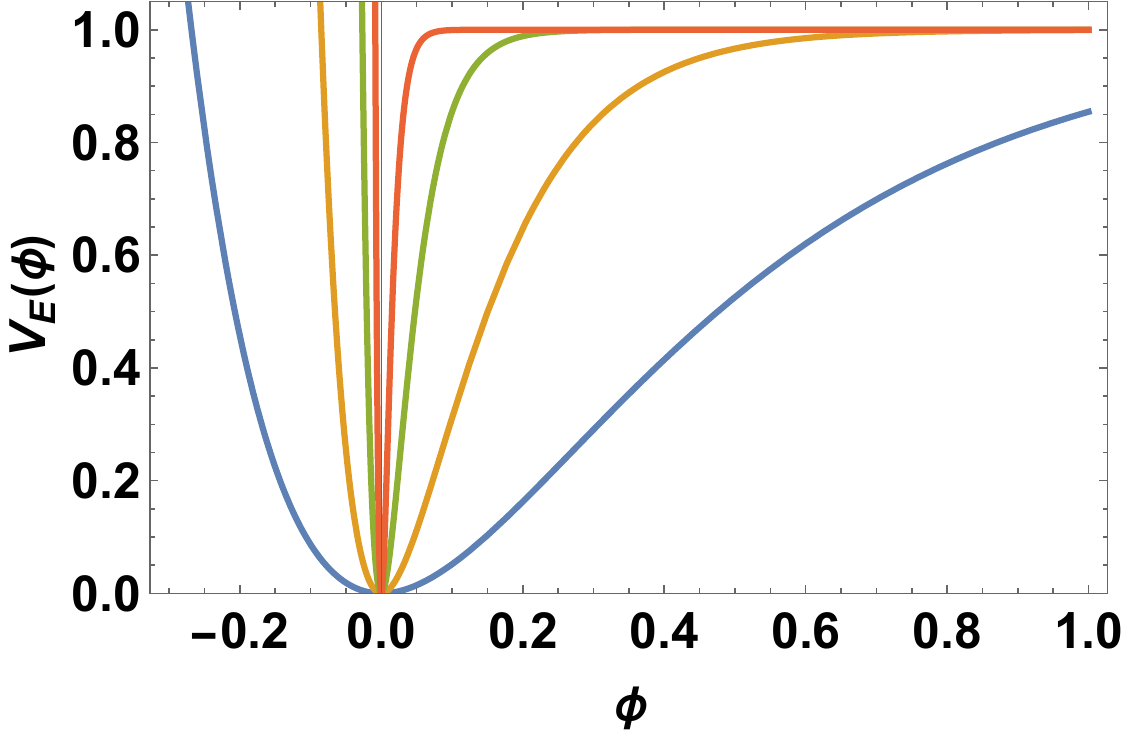}
        \label{fig:potentialsE}
    }
    \caption{Normalized $\alpha$-attractor potentials for \textbf{(a)} T-model and \textbf{(b)} E-model plotted for several values of $\alpha$: $10^{-1}$ in blue, $10^{-2}$ in orange, $10^{-3}$ in green, and $10^{-4}$ in red.}
    \label{fig:potentials}
\end{figure}

During preheating, oscillations of the scalar field at the bottom of the potential induce parametric instabilities in the perturbations, causing their growth. For potentials that approximate a parabola, the amplification of the Mukhanov–Sasaki (MS) variable is approximately given by $v_{\bm{k}} \sim a$, where $a$ is the scale factor of the universe. However, for $\alpha$-attractor potentials with small $\alpha$ (see Fig.~\ref{fig:potentials}), the potential strongly deviates from a quadratic form, and the amplification can be significantly stronger \cite{self-resonance}, affecting a wider range of scales. In this case, the parametric instability is known as self-resonance. This behavior and its implications are analyzed in detail in Sec.~\ref{sec:evol-perts}.

The evolution of perturbations during preheating is crucial for understanding PBH characterization, typically described in terms of their abundance and mass. The former is commonly obtained using the mass fraction $\beta(k)$, which quantifies the fraction of matter that collapses into PBHs. To calculate it, we employ two distinct approaches: {i)} the {altered} Press–Schechter (PS) formalism proposed in \cite{Martin:2019nuw} {which is different from the standard PS formalism known in the context of galaxies and clusters \cite{Press:1973} (see Appendix.~\ref{app:appendixA})}. {Worth noting that the altered PS formalism employs small values of threshold for an overdensity that can collapse, which is closely associated with what is commonly used in equation of state dependent collapse scenarios \cite{Harada:2013}}. {ii)} The Khlopov–Polnarev {(KP)} formalism \cite{Khlopov:1980,Khlopov:1981,Khlopov:1982}, is more appropriate for matter-dominated epochs, {which is not threshold dependent, but it distinctly incorporates non-spherical effects and pancake collapse of the density perturbations}. 

For estimating PBH masses, we adopt the standard approach of considering that the mass is roughly given by the horizon mass at the time the perturbation re-enters the horizon, although with some caveats. A comprehensive discussion of PBH characterization, including their abundance and mass distribution, is provided in Sec.~\ref{sec:PBH-characterization}. Finally, the conclusions of this study are summarized in Sec.~\ref{sec:conclusions}. {Furthermore, we also comment on the PBH formation analysis using the compaction function formalism in Appendix.~\ref{app:AppendixB} in the context of the preheating era, whose detailed study is, however, deferred for future investigations.}


\section{Constraints on evaporating primordial black holes}\label{sec:constraints}

\begin{figure*}
    \centering
    \includegraphics[width=0.75\linewidth]{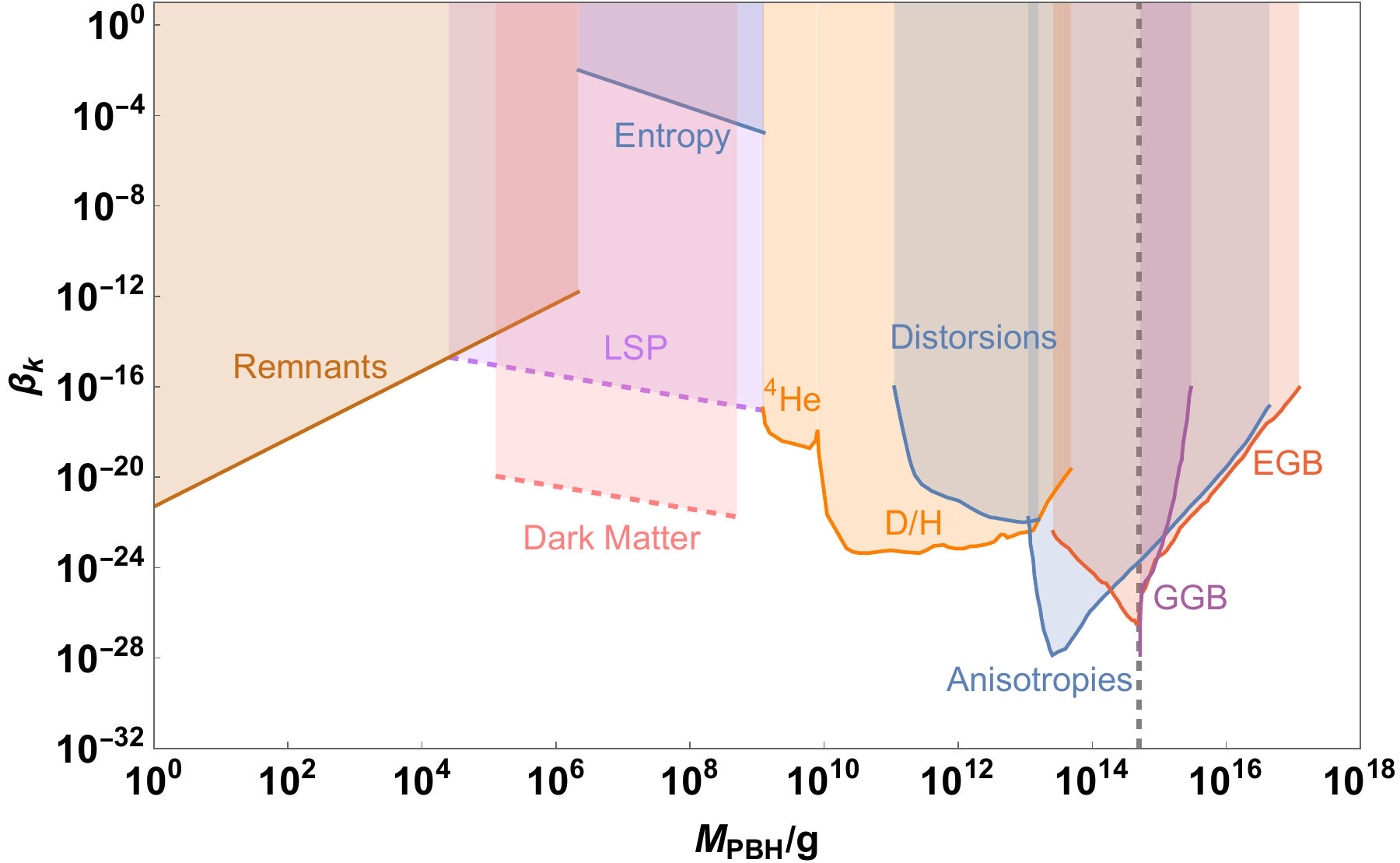}
    \caption{Constraints on the mass fraction of evaporating PBH from Planck remnants (brown), LSP (magenta) CMB effects (entropy, distortions, and anisotropies, in blue), BBN ($^4$He and D/H in orange), and $\gamma$-ray backgrounds (EGB in red and GGB in purple). The gray dashed vertical line marks the critical mass $M_{\text{crit}}$. The figure is produced from the data presented in \cite{Carr:2020,Gondolo:2020uqv} and references therein. We remark that these are just the strongest bounds based on analytical estimations and refer the interested reader to Fig.~4 of \cite{Carr:2020} for the full picture. For comparison, we also show the mass fraction of PBH obtained from the results of this work for an E-model and following the Khlopov-Polnarev (cyan) and Press-Schechter (green) formalism for $10$ e-folds of preheating and $\alpha=10^{-3}$. See the text below for details.}
    \label{fig:constraints}
\end{figure*}

Assuming the validity of Hawking radiation, one can impose some constraints on evaporating PBH. These constraints, summarized in Fig.~\ref{fig:constraints}, fall into five main categories.

\paragraph{Planck Remnants:}

The usual assumption is that the evaporation of a PBH proceeds until it vanishes \cite{Hawking:1974rv}. Based on the uncertainty principle \cite{Markov:1984xd,MacGibbon:1987my}, the information loss paradox \cite{Hawking:1974rv,Hawking:1976ra,Chen:2014jwq}, and some quantum gravity setups \cite{Aharonov:1987tp,Barrow:1992hq}, it was proposed that the black holes could stop evaporating at the Planck scales. Although we do not know what form of quantum gravity corrections should take
as one approaches the Planck mass, black hole remnant is an interesting hypothesis which we can test with observations  \cite{Carr:1994ar}\footnote{Note that the argument in favor of black hole remnant, though said to be motivated from quantum gravity theories, emerges from multiple ad-hoc assumptions to resolve information paradox. Alternatively, recent investigations of a unitary formulation of quantum field theory in curved spacetime in consideration with gravitational backreaction effects offer a potential resolution to the information paradox at the foundational level \cite{Kumar:2023hbj,Kumar:2024ahu}.}. In the case of PBH with $M_{\text{PBH}}\lesssim10^6g$, the rapid evaporation process could imply a significant population of remnants in the early universe. Still, their abundance must be limited as their density does not exceed the critical density of the universe \cite{Carr:1994ar}.

\paragraph{{Particle production from Hawking evaporation:}}

The evaporation of PBH could produce any particle predicted by theories beyond the standard model of particle physics. In particular, the evaporation of PBH may produce the lightest supersymmetric particles (LSP), predicted in supersymmetry and supergravity models, which are stable and may contribute to the dark matter. This affects PBH with masses $M_\text{PBH}\lesssim10^{11}(m_\text{LSP}/100\,\text{GeV})^{-1}$g \cite{Lemoine:2000sq}. This bound depends on the mass of the LSP particle and, therefore, can vary. In Fig.~\ref{fig:constraints} it is plotted for $m_{\text{LSP}}=100\,\text{GeV}$ in magenta color. Since in this work, we consider $\alpha$-attractor models that belong to the class of supergravity theories, we find it interesting to consider this constraint. {In \cite{Gondolo:2020uqv} it is considered that the PBH produce dark matter (DM) particles during their evaporation. Constraints are obtained based on the actual abundance of DM. We consider this scenario in Fig.~\ref{fig:constraints} for the case of Hawking temperature ($T_{\rm BH})$ of the PBH $T_{\text{BH}}>m_{\chi}$ and for DM particle mass of $m_{\chi}=10^{8}$\,GeV. These constraints are shown in dashed lines as they depend on the mass of the particle emitted.}

\paragraph{Cosmic Microwave Background (CMB):}

If PBH emit high-energy photons during evaporation, these would contribute to the photon-to-baryon, as first pointed out in \cite{Zeldovich}. Particularly, the photons from evaporating black holes of $M_{\text{PBH}}\lesssim10^9g$ would have enough time to thermalize, contributing to the photon-to-baryon ratio and thus increasing the entropy. However, this constraint is very weak, as can be seen in Fig.~\ref{fig:constraints}. On the other hand, as also pointed out in \cite{Zeldovich}, for $10^{11}g\lesssim M_{\text{PBH}}\lesssim10^{13}g$, the photons are emitted after the freeze-out of double-Compton scattering, which implies that the CMB photons develop a non-zero chemical potential, leading to a $\mu$-distortion. Another constraint related to CMB is the damping of small-scale anisotropies. If, during the evaporation process, the PBH also emit high-energy electrons, these can scatter with the CMB photons, contributing to the temperature anisotropy power spectrum. This constraint affects PBH in the range $2.5\times10^{13}g\lesssim M_{\text{PBH}}\lesssim2.4\times10^{14}g$ \cite{Acharya:2020jbv}, and is particularly strong. All the constraints related to CMB effects are shown in blue in Fig.~\ref{fig:constraints}.

\paragraph{Big-bang nucleosynthesis (BBN):}

The PBH of masses $10^9g\lesssim M_{\text{PBH}}\lesssim10^{13}g$ evaporate around the time of BBN. If during this process they inject high-energy particles into the plasma, these would modify the standard BBN scenario in two ways \cite{Carr:2009jm,Josan:2009qn}: (1) high-energy mesons and antinucleons modify the neutron-to-proton ratio, (2) high-energy hadrons and photons dissociate light elements such as $^4$He, thus increasing the abundance of others like D, T, $^3$He, $^6$Li and $^7$Li. These constraints are shown in orange in Fig.~\ref{fig:constraints} for the $^4$He and D cases.

\paragraph{$\gamma$-ray backgrounds:}

If the PBH with masses near $M_{\text{crit}}$ are evaporating today, then they can contribute to the extragalactic $\gamma$-ray background (EGB), as first pointed out by Page and Hawking in 1976 \cite{Page:1976}. They used observations of the EGB to constrain the mean number density of PBH evaporating at the present time, shown in red in Fig.~\ref{fig:constraints}. Also, if PBH are clustered within our galactic halo, then one would expect an anisotropic galactic $\gamma$-ray background (GGB), separable from the EGB. This constraint is shown in purple. Since these constraints are obtained through direct observations, as in the anisotropies case, they are very strong. Actually, one of the strongest constraints in the entire range of PBH masses, see for instance \cite{Carr:2020xqk,Carr:2020,Carr:2021bzv}.


\section{PBH formation during self-resonant preheating}\label{sec:pbh-formation}

\begin{figure}[htbp]
    \centering
    \subfigure[]{%
    \includegraphics[width=0.95\linewidth]{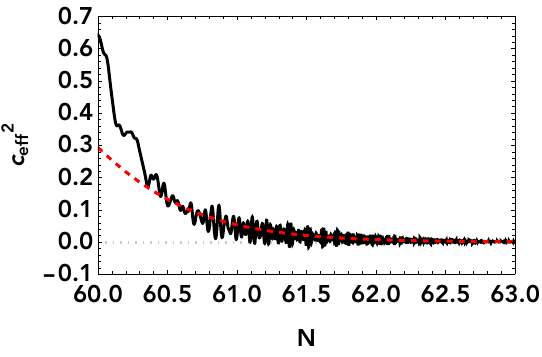}
        \label{fig:sound-speed-E-I}
    }
    \hfill
    \subfigure[]{%
        \includegraphics[width=0.95\linewidth]{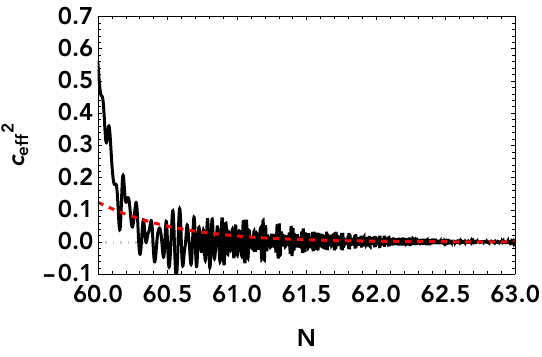}
        \label{fig:sound-speed-E-II}
    }
    \caption{{Numerically averaged (black) and analytical (dashed red) effective sound speeds, eqns.~\eqref{eq:speed-of-sound-numerical} and \eqref{eq:speed-of-sound-analytical}, respectively, for the wavenumber corresponding to the maximum amplification for an E-model with: \textbf{a)} $\alpha=10^{-2}$ and \textbf{b)} $\alpha=10^{-3}$.}}
    \label{fig:sound-speed}
\end{figure}

In this section, we describe how scalar perturbations can collapse to form PBH from instabilities produced by self-resonance effects during a matter-dominated phase, such as preheating. First, the range of modes affected by the self-resonance is what we will call \textit{instability band} (IB), whose upper limit is always determined by the Hubble radius $R_H=H^{-1}$, where $H(t)=\dot a/a$ is the Hubble rate, and a dot means derivation with respect to cosmic time. In contrast, the lower limit depends on the specific form of the potential. This last is referred to as the \textit{instability scale}  $l_{\text{inst}}$, which for a sufficiently long preheating ($>\mathcal{O}(1)$ e-folds) is given by \cite{self-resonance}
\begin{equation}
    l_{\text{inst}}^{-1}=\frac1M\left[\frac{2}{3\phi_{\textit{end}}}\left(\frac{a}{a_{\textit{end}}}\right)^3\right]^{1/4},
\end{equation}
where the suffix ``$_{\text{end}}$'' means evaluation at the end of inflation. The value of the field at the end of inflation for the T- and E-models is obtained using slow-roll conditions and is given, respectively, by \cite{Shafi:2024jig}
\begin{subequations}\label{eq:phi-end}
\begin{gather}
    \phi_{\text{end}}^T\approx \Mpl\sqrt{\frac{3\alpha}{2}}\arcsinh{\left(\frac2{\sqrt{3\alpha}}\right)},\\\phi_{\text{end}}^E\approx\Mpl\sqrt{\frac{3\alpha}{2}}\log{\left(\frac2{\sqrt{3\alpha}}+1\right)}.
\end{gather}
\end{subequations}
The characteristic scale $l_{\text{inst}}$ defines the IB as the range of modes satisfying:
\begin{equation}\label{eq:IB-criterion}
        l_{\text{inst}}<\frac{a}{k}<R_H(t_{\text{rh}}),
\end{equation}
This is what we call the IB criterion. Perturbations whose comoving wavenumber $k$ is such that $a/k<l_{\text{inst}}$ during preheating, decay and therefore do not favor the collapse. Here, the suffix ``$_{\text{rh}}$'' means evaluation at the end of preheating. Using \eqref{eq:phi-end}, we observe that as $\alpha$ decreases, so does the instability scale $l_{\text{inst}}$, and thus the IB gets broadened, increasing the number of modes that are affected by the self-resonance, and potentially the abundance of PBH. Next, we consider the Jeans' length criterion, which states that the size of a perturbation\footnote{The typical size of a perturbation is taken as half the physical wavelength of the perturbation, which is given by {$\lambda_{\text{phys}}=\frac{a\lambda}{2}=\frac{a\pi}{k}$}} must be larger than the Jeans' length so that pressure cannot counteract gravity and be able to collapse while remaining smaller than the Hubble radius in order to preserve causality. This condition is expressed as:
\begin{equation} \label{eq:jeans-length}
    \frac{R_J}{\pi}<\frac{a}{k}<R_H,
\end{equation}
where $R_J\simeq\sqrt{2c_{\text{eff}}^2/3}\,R_H$ is the Jeans' length \cite{Harada:2013,Niemeyer:2019aqm,Reis:2003fs} and $c_{\text{eff}}^2$ denotes the effective speed of sound. This criterion imposes an upper limit on the wave numbers $k$ that can potentially collapse, which competes with $l_{\text{inst}}^{-1}$. The lower bound for the IB is chosen as the larger of these two scales, ensuring both criteria are met. For a perfect fluid with an equation of state $w$, the speed of sound is given by $c_s^2=w$. However, for scalar fields, this is derived differently \cite{del-Corral:2023,Cembranos:2015oya,Hertzberg:2014iza}, and perturbation theory is required. In this case, we have an effective (or averaged) sound speed {$c_{\text{eff}}^2$} defined in general as the ratio of pressure to density perturbations{. By using the adiabatic expansion described in \cite{Cembranos:2015oya}, one obtains the following expression}
{\begin{equation}\label{eq:speed-of-sound-numerical}
\begin{split}
    c_{\text{eff}}^2(k)\equiv\frac{\langle\delta p_{\bm k}\rangle}{\langle\delta\rho_{\bm k}\rangle}\simeq\frac{\langle\frac{k^2}{a^2}\phi-V'(\phi)+V''(\phi)\phi\rangle}{\langle\frac{k^2}{a^2}\phi+3V'(\phi)+V''(\phi)\phi\rangle},
\end{split}
\end{equation}}
where $\langle\dots\rangle$ denotes averaging over a time interval larger than the oscillation period but small compared to the Hubble time \cite{Cembranos:2015oya}. {This ensures that the fast dynamics of the scalar field are properly captured while ignoring the ``slow'' cosmological changes during the averaging.} For inflationary potentials whose expansion in powers of $\phi$ can be written as
\begin{equation}
V(\phi)\sim \frac{M^2}2\phi^2+\frac{\lambda_3}{3}\phi^3+\frac{\lambda}{4}\phi^4+\dots,
\end{equation}
as it is the case of the $\alpha$-attractors, the {effective} speed of sound can be approximated analytically as
\begin{equation}\label{eq:speed-of-sound-analytical}
    {c_{\text{eff}}^2}\simeq\frac{k^2+2\lambda a^2\langle\phi^2\rangle}{k^2+4M^2a^2+6\lambda a^2\langle\phi^2\rangle}.
\end{equation}
Here, we have used that the average of even powers of the scalar field vanishes due to its oscillatory behavior \cite{Cembranos:2015oya}. For this reason, ${c_{\text{eff}}^2}$ is independent of $\lambda_3$. The parameter $\lambda$ for the T- and E-models is given, respectively, by
\begin{equation}
    \lambda_T=-\frac{2M^2}{9\alpha},\qquad\lambda_E=\frac{7M^2}{9\alpha}.
\end{equation}
In the high-$k$ limit, both density and pressure perturbations average to zero, leading to ${c_{\text{eff}}^2}=1$, which holds both for the full expression \eqref{eq:speed-of-sound-numerical} and the analytical approximation \eqref{eq:speed-of-sound-analytical}. This implies that a sub-horizon mode must initially be in the order of the Hubble radius to collapse, which, by definition, is impossible. To validate this approximation, Fig.~\ref{fig:sound-speed} compares the full numerical calculation of ${c_{\text{eff}}^2}$ from \eqref{eq:speed-of-sound-numerical} (black) with the analytical result \eqref{eq:speed-of-sound-analytical} (dashed red) for {a E-model with $\alpha=10^{-2}$ and $\alpha=10^{-3}$ for the wavenumber where the maximum amplification occurs. One can check that, by the end of inflation, the speed of sound is nonzero, but as preheating proceeds, it decreases towards zero (on average). At this point, the Jeans' length is sufficiently small to allow the mode to collapse.}

Lastly, a perturbation will collapse into a PBH if the density contrast, defined as $\delta_{\bm{k}} = \frac{\delta\rho_{\bm{k}}}{\rho}$, exceeds a threshold value $\delta_c^{\text{min}}$. Here, $\rho$ and $\delta\rho_{\bm{k}}$ represent the background density and the density perturbation, respectively. The simplest estimate for $\delta_c$ arises from Carr's original analysis \cite{Carr:1975,Escriva:2021}, which gives $\delta_c \simeq w$. However, instead of using Carr's criterion (suitable for a fluid description), we compute the threshold using time constraints \cite{del-Corral:2023,Martin:2019nuw,Jerome:2020b,Jedamzik:2010dq}. Specifically, the time a perturbation $\delta_{\bm k}$ needs to collapse into a PBH can be computed using the top-hat spherical collapse model of the massive scalar field in the Einstein de Sitter universe, which is given by \cite{Goncalves:2000nz,Martin:2019nuw}
\begin{equation}\label{eq:time-of-collapse}
    \Delta t_{\text{coll}}=\frac{\pi}{H[t_{\text{bc}}(k)]\delta_{\bm k}^{3/2}[t_{\text{bc}}(k)]},
\end{equation}
where $t_{\text{bc}}(k)$ is the cosmic time when the mode enters the IB. Eqn.~\eqref{eq:time-of-collapse} effectively estimates the time needed for a perturbation to grow to $\mathcal{O}(1)$, assuming that 
$\delta_{\bm k}\sim a$, which allows us to define the threshold as follows \cite{Martin:2019nuw}
\begin{equation}\label{eq:formal-threshold}
    \delta_c^{\text{min}}(k)=\left(\frac{3\pi}2\right)^{2/3}\left[e^{3(N_\text{rh}-N_\text{bc}(k))/2}-1\right]^{-2/3},
\end{equation}
where $N_\text{rh}$ is the number of e-folds at which preheating ends and $N_\text{bc}(k)$ is the number of e-folds at which a perturbation enters the IB. However, this assumption holds when self-resonance is weak. As we have seen, this assumption breaks down for small values of $\alpha$ \cite{self-resonance}. Thus, instead of using \eqref{eq:formal-threshold} we numerically compute, for each value of $\alpha$ and $k$, the minimum value of the density perturbation that reaches $\mathcal{O}(1)$. This minimum value serves as the threshold $\delta_c^{\text{min}}(k)$, which is now $k$-dependent by construction. Thus, if $\delta_{\bm k}$ is larger than its corresponding minimum value $\delta_c^{\text{min}}(k)$, then the mode has enough time to reach $\mathcal{O}(1)$ and potentially collapse. In essence, this implies that due to the self-resonance effects, the threshold is expected to decrease as $\alpha$ decreases, since the perturbations grow faster and the minimum value that each mode needs to collapse decreases, which ultimately enhances the production of PBH.

In summary, since we are working with small-scale instabilities, we use three criteria to select the modes that can collapse into PBH:
\begin{itemize}
    \item \textit{Instability Band Criterion:} Modes must lie within the IB to be affected by the instabilities that favor the collapse, satisfying \eqref{eq:IB-criterion}.
    \item \textit{Jeans' Length Criterion:} Modes must satisfy \eqref{eq:jeans-length} to ensure that during the collapse, the pressure does not counteract gravity. The speed of sound is given by \eqref{eq:speed-of-sound-analytical}.
    \item \textit{Density Contrast Criterion:} The density contrast $\delta_{\bm{k}}$ must exceed the threshold $\delta_c^{\text{min}}(k)$, to allow sufficient time for PBH formation.
\end{itemize}

\begin{figure}[htbp]
    \centering
    \includegraphics[trim = 20mm 90mm 5mm 0mm,width=0.95\linewidth]{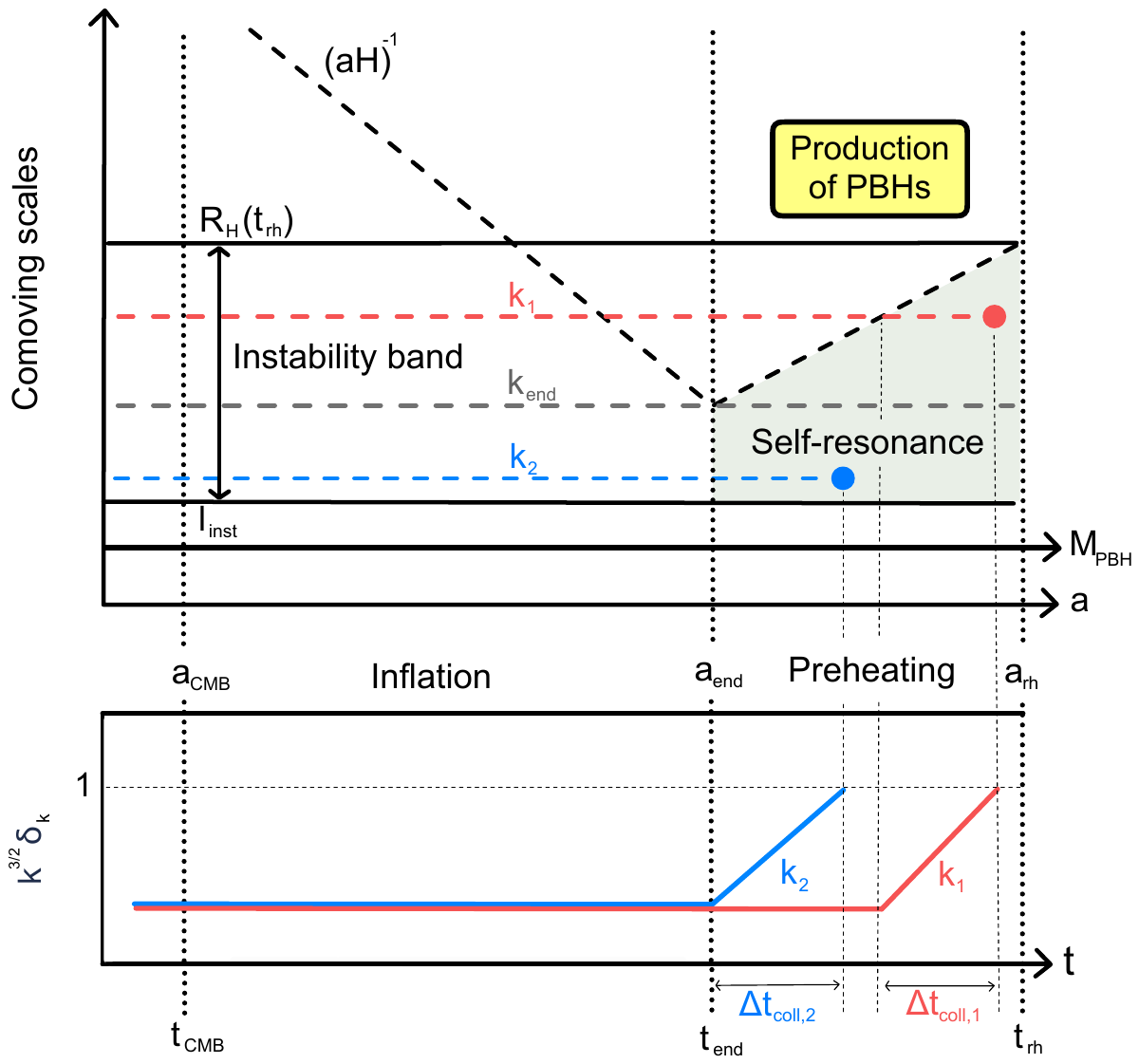}
    \caption{Schematic representation of the collapse of perturbations during self-resonant preheating, following the approach described in this work. The upper panel shows the evolution of the comoving scales involved. The lower panel is a representation of the evolution of the density contrast for two particular modes $k_1$ and $k_2$. See text for details.}
    \label{fig:summary}
\end{figure}

The situation is summarized in Fig.~\ref{fig:summary}. The upper panel shows the evolution of some representative comoving scales. Particularly, the IB corresponds to the range of scales between $R_H(t_{\text{rh}})$ and $l_{\text{inst}}$, and amplification is assumed to occur as soon as a mode enters it. It is labeled as the self-resonance instability region (green-shaded area) in Fig.~\ref{fig:summary}. The time $t_{\text{rh}}$ marks the end of the preheating and the beginning of reheating. Also shown as an example are the two scales $k_1$ (red) and $k_2$ (blue). The former enters the IB after the end of inflation, whereas the latter is already inside by the end of inflation. Once inside the IB, their density contrast (plotted in the lower panel) grows until it reaches non-linearity in a time $\Delta t_{\text{coll}}$. If the time they spent growing inside the IB is long enough so that the collapse occurs during preheating, \textit{i.e.} $\delta_{\bm k}>\delta_c^{\text{min}}(k)$, then we say that the mode has collapsed into a PBH (represented by a circle), and its mass is given by the mass of the horizon at the moment they enter the IB, as it is usually assumed \cite{Martin:2019nuw}. In this case, although both modes have enough time to collapse, the thresholds are different in each case. Specifically, $\delta_c^{\text{min}}(k_2)>\delta_c^{\text{min}}(k_1)$, since the minimum value of the density perturbation the mode $k_2$ needs to reach $\mathcal{O}(1)$ is lower than the case with $k_1$ because $k_2$ spends more time in the IB and thus can grow higher than $k_1$. For simplicity, we do not show the Jeans' length. However, it can be shown to be of the order of the instability scale. 
We acknowledge that the reader may be concerned about the status of the modes whose wavenumber $k$ is such that ${\frac{a}{l_{\text{inst}}}}>k>k_{\text{end}}$ and therefore these have not exited the horizon during inflation{, but they fall within the IB}. However, we showed in [9] that all the modes behave identically once inside the IB, {with no decay whatsoever, }which motivates us to include them in our analysis and explore their consequences, bearing in mind their different nature. {However, the modes $k>\frac{a}{l_{\text{inst}}}$ do present a suppression, since the curvature perturbation decays for these modes as one would typically expect. Therefore, the production of PBHs within these modes is not taken into account in our analysis as they lie outside the IB and therefore are not.}


\section{Evolution of perturbations}\label{sec:evol-perts}

\begin{figure}[htbp]
    \centering
    \subfigure[]{%
    \includegraphics[width=0.95\linewidth]{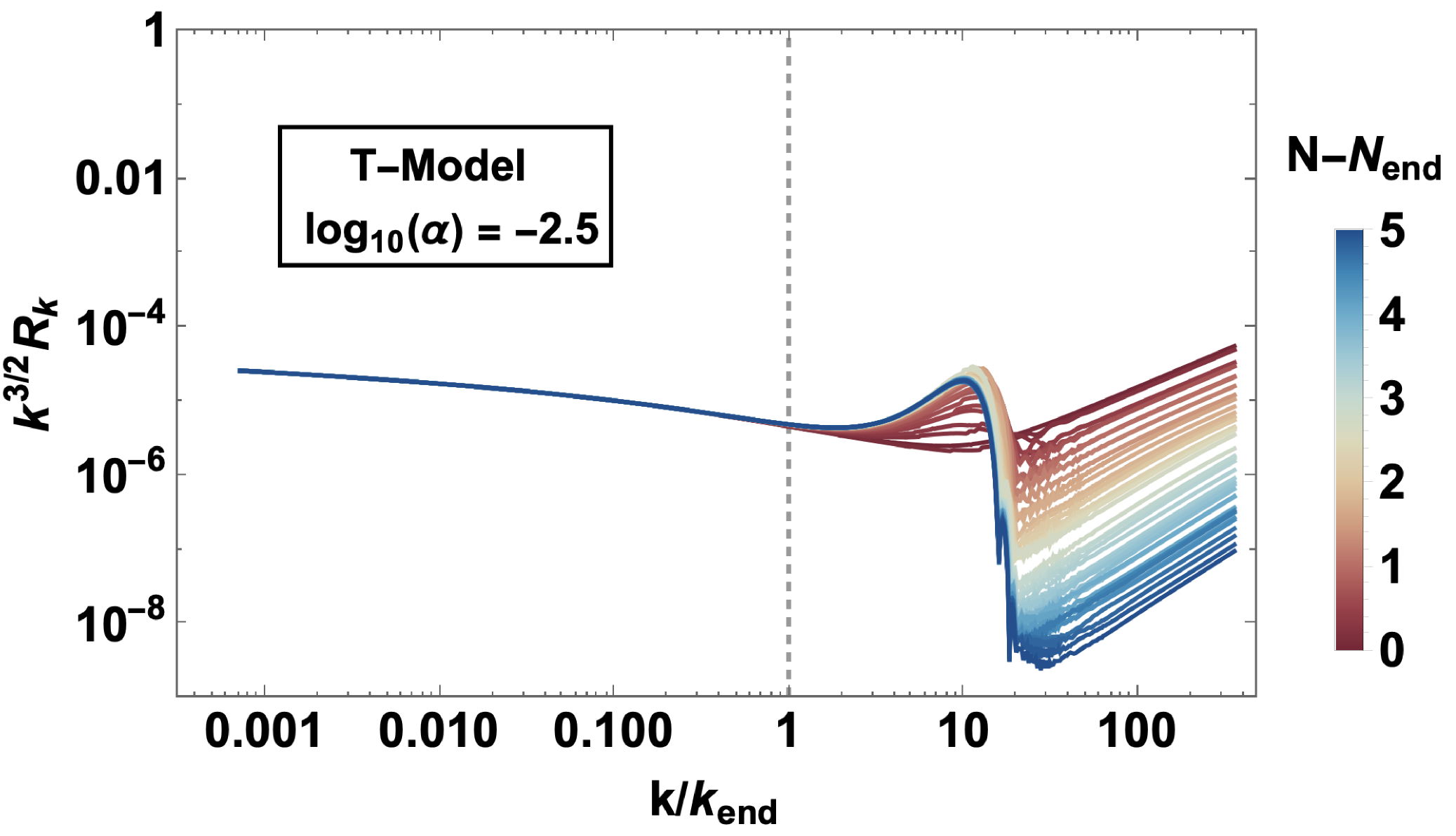}
        \label{fig:curvature-T25}
    }
    \hfill
    \subfigure[]{%
    \includegraphics[width=0.95\linewidth]{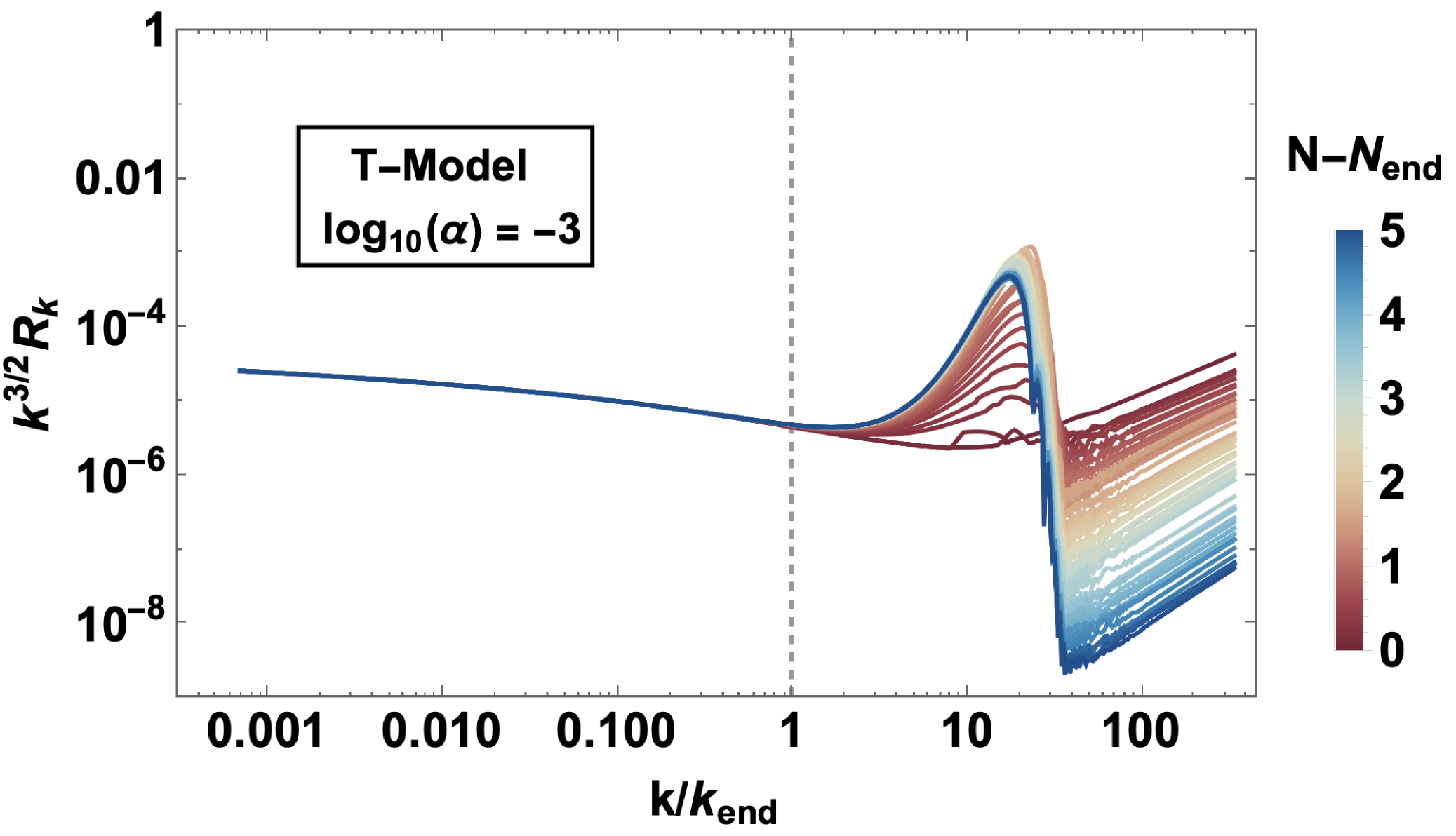}
        \label{fig:curvature-T3}
    }
    \hfill
    \subfigure[]{%
        \includegraphics[width=0.95\linewidth]{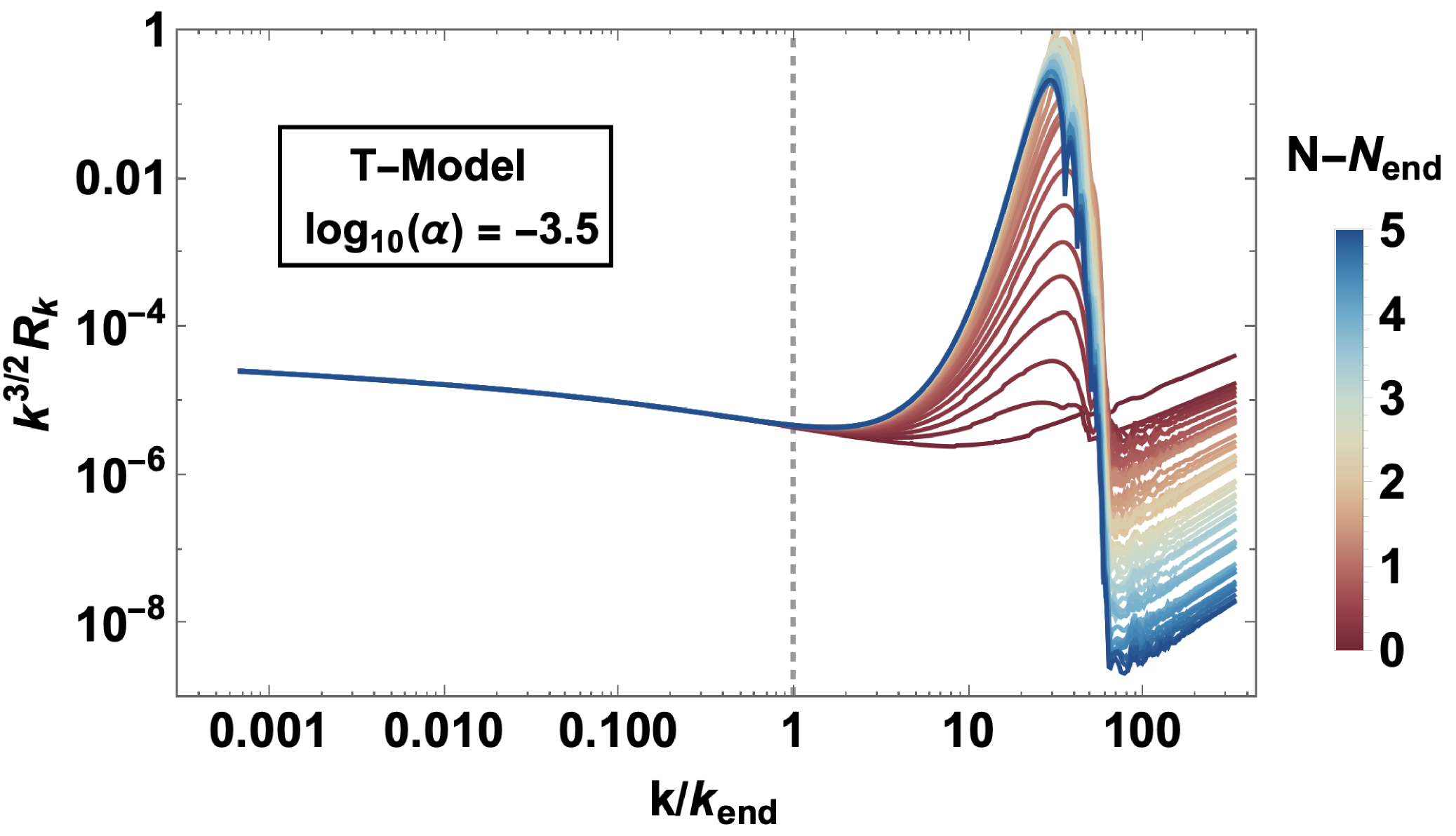}
        \label{fig:curvature-T35}
    }
    \caption{{Normalized curvature perturbations, $k^{3/2}\mathcal{R}_{\bm k}$, for a T-model with several values of $\alpha$. The time evaluation is made from the end of inflation ($N_{\text{end}}=60$) to 5 e-folds after, in steps of 0.1 e-folds. The vertical grey line marks the scale that exits the horizon at the end of inflation for each model, $k_{\text{end}}$}}
    \label{fig:curvature}
\end{figure}

To describe the cosmological evolution of a scalar field minimally coupled to gravity we use the following Lagrangian
\begin{equation}\label{eq:lagrangian}
    \mathcal{L}=-\frac12g^{\mu\nu}\partial_{\mu}\phi\partial_{\nu}\phi-V_{T,E}(\phi),
\end{equation}
where $g_{\mu\nu}$ is the background space-time metric. To study the scalar perturbations, we introduce a fluctuation in the scalar field as $\phi\rightarrow\phi+\delta\phi$, which sources scalar perturbations $\Phi$ and $\Psi$ in the background metric, given by the following perturbed FLRW line element
\begin{equation}
    \dd s^2=a^2[-(1+2\Phi)\dd \eta^2+(1-2\Psi)\delta_{ij}\dd x^i\dd x^j],
\end{equation}
where $\delta_{ij}$ is the Kronecker delta, $\eta$ conformal time, related to cosmic time $t$ through the relation $\dd\eta=\dd t/a$, and we have chosen to work in Newtonian gauge. Considering that the matter content is described by the scalar field $\phi$, then the Einstein equations in the perturbed FLRW metric lead to a simple equation when written in conformal time. This equation is known as the Mukhanov-Sasaki equation, which in Fourier space is given by
\begin{equation}\label{eq:MS-tau}
    v''_{\bm k}(\eta)+\left[k^2-\frac{\left(a\sqrt{\epsilon_H}\right)''}{a\sqrt{\epsilon_H}}\right]v_{\bm k}(\eta)=0,
\end{equation}
where {an overprime} means derivation with respect to conformal time, $\epsilon_H=1/2(\phi'/\mathcal H)^2$ is the first slow-roll parameter and $\mathcal H=a'/a^2$ is the conformal Hubble rate. Also, a suffix ``$_{\bm k}$'' indicates we are working in Fourier space, where $\bm{k}$ is the wavenumber vector and $k=|\bm{k}|$ its modulus. The MS variable is defined in terms of a combination of the field and metric perturbations, \textit{i.e.}, $v_{\bm k}=a\left[\delta\phi_{\bm k}+\phi'\Phi_{\bm k}/\mathcal{H}\right]$ (see \cite{Mukhanov:1990me,Baumann:2009ds} for further details). This equation is typically solved in conformal time during inflation. However, during preheating $\epsilon_H$ oscillates around zero, which produces a singular behavior in \eqref{eq:MS-tau}. To circumvent this and solve for $v_{\bm{k}}$ one can change it to cosmic time $t$. Then, as shown in \cite{self-resonance}, during a self-resonant phase, one can perturbatively solve the evolution equation of the inflation field as a series expansion of a small parameter $\beta$ as follows 
\begin{equation}
\phi(\tau)=\sum_{n=1}^{\infty}\beta^n\phi_n(\tau),
\end{equation}
where $\tau=M\sqrt{1-\beta^2}\,t$, so that the MS equation \eqref{eq:MS-tau} can be transformed into the following Hill equation 
\begin{equation}\label{eq:hill}
    \frac{\dd^2 \tilde v_{\bm k}}{\dd\tau^2}+\left[A_k+\sum_{n=1}^{\infty}\Big(q_n(\tau)\cos(n\tau)+p_n(\tau)\sin(n\tau)\Big)\right]\tilde v_{\bm k}=0,
\end{equation}
where $\tilde{v}_{\bm k}=a^{1/2}v_{\bm k}$. The functions $A_k$, $q_n$, and $p_n$ depend upon the parameters of the model and the potential has been expanded in powers of $\phi$. Using the Floquet theorem, the  rescaled MS variable in \eqref{eq:hill} is found to evolve as $\tilde v_{\bm k}\sim\exp{\int\mu_k\dd\tau}$, where $\mu_k$ are the so-called Floquet exponents, which in turn depend on the functions $A_k$, $q_n$ and $p_n$. For $\Re(\mu_k)>0$, we have exponential amplification, and thus the mode is said to be unstable. This translates into a strong amplification for a particular range of modes. To show this, we relate the MS variable with the curvature perturbation $\mathcal R_{\bm k}$ as follows
\begin{equation}\label{eq:curvature}
\mathcal{R}_{\bm k}=\frac{v_{\bm k}}{\sqrt{2\epsilon_H}a}=\frac{\tilde v_{\bm k}}{\sqrt{2\epsilon_Ha^{3}}}.
\end{equation}
Fig.~\ref{fig:curvature} shows the evaluation of the power spectrum of the curvature perturbations at different times after the end of inflation and for several values of $\alpha$. The underlying model is a T-model, and the evaluations are made in steps of 0.1 efolds from the end of inflation $N_{\text{end}}=60$, to 5 efolds after. Notably, as $\alpha$ decreases, the curvature perturbation $\mathcal{R}_{\bm k}$ amplifies faster and in a larger amount after the end of inflation, caused by the self-resonance effect. Also, as a result of this decrease in $\alpha$, the IB broadens, and more modes fall within and get amplified.

We numerically and analytically showed in \cite{del-Corral:2023} that during preheating and for the modes inside the IB, both $\mathcal{R}_{\bm k}$ and Bardeen potential $\Phi_{\bm k}$ behave in the same way with amplitudes differing in a numerical factor of order $\mathcal{O}(1)$. Using this, one can operate with the perturbed Einstein equations to arrive at the following expression for the density contrast in terms of the curvature perturbation
\begin{equation}\label{eq:density-pert}
    |\delta_{\bm k}|\sim\frac25\left(\frac{k^2}{a^2H^2}+3\right)|\mathcal{R}_{\bm k}|.
\end{equation}
Since the first term inside the parenthesis grows as $\sim a$ during preheating and $\mathcal{R}_{\bm k}$ is at least constant but never decaying inside the IB, we see that density perturbations grow inside the IB as long as $k^2/(a^2H^2)>3$, and the growth depends on the Floquet exponents $\mu_k$ which ultimately is given by the value of $\alpha$. {Let us see this in more detail. The expressions for $\mu_k$ are, in general, complicated. However, for a quadratic potential $V\sim\phi^2$, it can be shown that the dominant term of the Floquet exponent goes as \cite{self-resonance}
\begin{equation}
    \mu_k\sim \mu_0\left(\frac{a_{\text{end}}}{a}\right)^{3/2},
\end{equation}
where $\mu_0$ is a constant of $\mathcal{O}(1)$. This implies that
\begin{equation}
    \tilde{v}_{\bm k}\sim e^{\int\mu_k\dd\tau}\sim e^{\int M(a_{\text{end}}/a)^{3/2}\dd t}\sim t\sim a^{3/2},
\end{equation}
where we have used the fact that during preheating $a\sim t^{2/3}$, $\tau=Mt$, and $t_{\text{end}}\sim M^{-1}$. Using this, the curvature perturbation in eqn.~\eqref{eq:curvature} is given by $\mathcal{R}_{\bm k}\sim \text{cte}$, which, according to eqn.~\eqref{eq:density-pert}, implies $\delta_{\bm k}\sim a$. This is the standard situation described in many preheating scenarios \cite{Jedamzik:2010dq,Jerome:2020b,Martin:2019nuw,Martin:2020fgl}, but it is only true for quadratic potentials. Only in those cases we can say that the density contrast grows linearly with the scale factor. For potentials deviating from the quadratic behaviour, the curvature perturbation is not constant anymore due to the self-resonance effects (as one can see from Fig.~\ref{fig:curvature}), and actually grows. Besides, the Floquet exponent gets an extra contribution that depends on $\alpha$ \cite{self-resonance} which is schematically given by
\begin{equation}
\mu_k\sim\mu_0\sqrt{\frac1{\alpha}\left(\frac{a_{\text{end}}}{a}\right)^6+\left(\frac{a_{\text{end}}}{a}\right)^{3}}.
\end{equation}
At early times, the first term dominates for small $\alpha$, which ultimately makes $\delta_{\bm k}$ to grow, at least, at a higher rate than linear with $a$. Then, as preheating goes on, this first term decays faster, and we recover the standard quadratic behaviour. Therefore, for potentials deviating from a quadratic shape, we can extract from the above analysis two conclusions. i) The density contrast does not grow linearly with the scale factor anymore, and at least it grows at a higher rate. If the growth is quadratic, cubic, or any other rate, it would require a detailed mathematical analysis far from the objectives of this work. ii) Due to the $1/\alpha$ factor on the expression for the Floquet exponent, the rate of growth will be higher as we decrease this parameter.}

It is important to note that both the field perturbations $\delta\phi_{\bm{k}}$ and the metric perturbations $\Phi_{\bm{k}}$ are included in the equations through the MS variable $v_{\bm{k}}$. Metric perturbations are often neglected in lattice simulations (see, e.g., \cite{Ballesteros:2024hhq}), where a suppression of field perturbation amplification is observed. However, as shown in \cite{self-resonance}, for small values of $\alpha$, these perturbations can become significantly amplified, justifying their inclusion in our analysis. Furthermore, we recall that our analysis includes fluctuations that never exit the horizon but become unstable during preheating. Beyond this semi-analytical approach, one can also employ fully non-linear general-relativistic field-theory simulations, as demonstrated in \cite{Aurrekoetxea:2023jwd}, where the formation of oscillons during preheating is investigated.


\section{Primordial black hole characterization}\label{sec:PBH-characterization}

Having understood the growth of density perturbations within the IB, we now proceed to characterize the PBH by calculating their mass fraction and corresponding masses. To compute the mass fraction, we employ the {(altered)} Press-Schechter (PS), and Khlopov-Polnarev (KP) formalisms.

{In the (altered) PS framework~\cite{Press:1973,Martin:2019nuw,Harada:2013}, one implements a statistical mapping from the Gaussian probability distribution of smoothed density fluctuations to the PBH mass spectrum by counting regions whose amplitude exceeds a critical collapse threshold. This prescription treats numerous independent, small-amplitude peaks within a Hubble volume and converts the statistical tail of a Gaussian field into a mass function through a normalization that compensates for cloud-in-cloud issues.  It is suitable when large fluctuations are relatively common and overlap, while each individual fluctuation remains weak. In contrast, the KP scenario~\cite{Khlopov:1980,Khlopov:1981,Khlopov:1982} addresses PBH formation during an early matter-dominated epoch with nonrelativistic matter. Their analysis extends Zel’dovich’s approximation by considering anisotropic collapse into flattened, sheet-like configurations and by estimating the probability that such nearly planar overdensities satisfy the conditions for black hole formation. In our case, however, the fluctuations generated during preheating in $\alpha$-attractor models do not fall into either category. They are neither the numerous small peaks envisioned by PS nor the isolated, weak overdensities evolving into thin pancakes as in KP. Instead, we find many large-amplitude fluctuations on scales comparable to or larger than about one tenth of the Hubble radius, a regime that lies outside the direct assumptions of both formalisms. For this reason, we employ the (altered) PS and KP prescriptions as opposite extremes to estimate the PBH mass function. The fact that they lead to results differing by many orders of magnitude shows the limitations of applying either framework directly to our situation and points to the need for a more suitable analysis of collapse in this regime.}


\subsection{Mass fraction}\label{sec:mass-fraction}

The mass fraction $\beta(k)$ is the main quantity used to characterize the abundance of PBH. It represents the fraction of the total mass that goes into PBH. It is also common to express it normalized to the fractional density of dark matter, $\Omega_{\text{DM}}$. In that case, it is typically represented by $f(M)$. We do not take into account for this work the dark matter in this work and thus restrict ourselves to $\beta(k)$.

\paragraph{Press-Schechter.}

In the context of PBH, the PS formalism is commonly used \cite{Harada:2013} to compute the mass fraction in a radiation-dominated universe, where the equation of state parameter is $w = 1/3${, which gives a collapse threshold of $\delta_c^{\text{rad}}\sim w\sim1/3$ \cite{Carr:1975}}. However, as discussed previously, we apply this formalism to the dust-like preheating stage, characterized by $w \sim 0$. {In Appendix~\ref{app:appendixA}, we show that in this situation one can apply the top-hat spherical collapse model, and the collapse threshold is given by $\delta_c^{\text{lin}}=1.686$.} The PS method assumes Gaussian statistics $P$ for the density perturbations, $P(\delta)$, and the variance is typically given by the power spectrum of density perturbations, $\sigma_k^2\simeq\mathcal{P}_\delta(k)$, where this last is defined by
\begin{equation}
    \mathcal{P}_\delta(k)=\frac{k^3}{2\pi^2}|\delta_{\bm k}|^2.
\end{equation}
The expression of $\beta(k)$ within this formalism is given by
\cite{Press:1973,Harada:2013}
{\begin{equation}\label{eq:PS-formalism} 
\begin{split}
\beta(k)
&= \frac{\dd\Omega_{\text{PBH}}(k)}{\dd \ln{M}}= 2 \int_{\delta_{c}^{\text{lin}}}^{\delta_{\mathrm{max}}} P(\delta) \dd\delta 
\\&\simeq\erfc\left[\frac{\delta_{c}^{\text{lin}}}{\sqrt{2}\sigma_k}\right]-\erfc\left[\frac{\delta_{\text{max}}}{\sqrt{2}\sigma_k}\right], 
\end{split}
\end{equation}}
where $\erfc$ is the complementary error function, and $\delta_{\text{max}}$ is the maximum threshold value, usually set to $\delta_{\text{max}}=1$ to avoid entering the non-linear regime. {The variance $\sigma_k$ should be evaluated at the moment the perturbations enter the IB, when their amplitudes are of order $\mathcal{O}(10^{-5})$, far from the threshold $\delta_c^{\text{lin}}=1.686$. This gives a difference of 5 orders of magnitude and, according to eqn.~\eqref{eq:PS-formalism}, a null production of PBHs, since $\beta(k)\simeq\erfc(10^5)\cong0$. This implies that the standard PS formalism is not useful in this case to estimate the production of PBHs.}

\paragraph{{Altered Press-Schechter.}}

{Following the strategy of \cite{Martin:2019nuw}, we compute an effective initial critical value $\delta_c^{\rm min}(k)$ for collapse into PBHs, and replace it with the lower bound in the integral \eqref{eq:PS-formalism} (from the PS formalism), which we now call altered-PS. This effective threshold is obtained just by taking into account that the linear density contrast grows as $\delta_{\bm k}\sim a$ and that the fluctuations take a time $\Delta t_{\text{coll}}$ (see eqns.~\eqref{eq:time-of-collapse} and \eqref{eq:formal-threshold}) to grow to the critical value $\delta_c^{\text{lin}}=1.686$ and collapse. Then, they substitute this scale-dependent threshold into the PS formula \eqref{eq:PS-formalism} to estimate $\beta(k)$. This, however, departs from the strict, original PS prescription \cite{Press:1973}, as shown in the previous subsection. To be more precise, one should label this effective initial value as $\delta_c^{\text{min}}(k)$ instead of $\delta_c^{\text{lin}}$, as it is the minimum value a perturbation needs to collapse, not the precise threshold amplitude. The mass fraction is now estimated as
\begin{equation}\label{eq:modified-PS}
    \beta(k)\simeq\erfc\left[\frac{\delta_{c}^{\text{min}}(k)}{\sqrt{2}\sigma_k}\right]-\erfc\left[\frac{\delta_{\text{max}}}{\sqrt{2}\sigma_k}\right].
\end{equation}}

\begin{figure}[htbp]
    \centering
    \subfigure[]{%
    \includegraphics[width=0.95\linewidth]{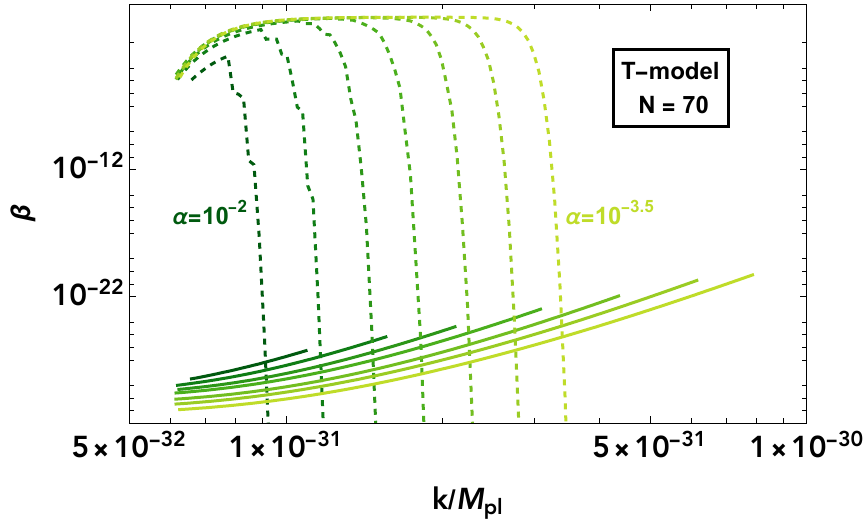}
        \label{fig:T-model-75}
    }
    \hfill
    \subfigure[]{%
        \includegraphics[width=0.95\linewidth]{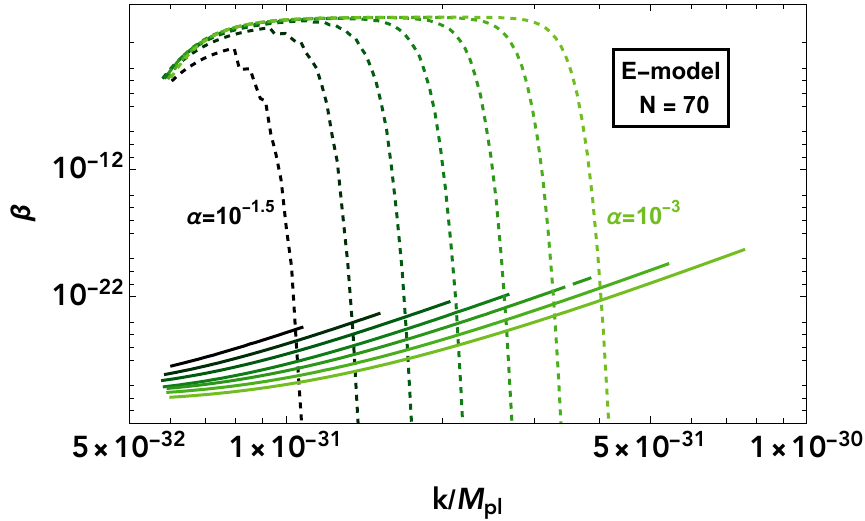}
        \label{fig:E-model-75}
    }
    \caption{Mass fraction for \textbf{(a)} T-model and \textbf{(b)} E-model using both KP \eqref{eq:KP-formula} and {altered-}PS \eqref{eq:modified-PS} formalisms in dashed and continuous, respectively. The evaluation is made 10 e-folds after the end of inflation and values of $\alpha$ go from $10^{-2}$ to $10^{-3.5}$ for the T-model and from $10^{-1.5}$ to $10^{-3}$ for the E-model, in both cases with incremental steps of $10^{0.25}$. The color code of each value of $\alpha$ is the same in both plots. The lower value for $\alpha$ in each model is taken as the lower bound on $\alpha$ obtained in \cite{del-Corral:2025fzz}.}
    \label{fig:beta}
\end{figure}

In Fig.~\ref{fig:beta} we plot the mass fraction $\beta(k)$ following the {altered}-PS formalism (dashed curves) as a function of $k$ for both T- and E-models and different values of $\alpha$. Only the modes that satisfy the three criteria defined in Sec.~\ref{sec:pbh-formation} are considered, with evaluations performed 10 e-folds after the end of inflation. 
The results reveal that as $\alpha$ decreases, the mass fraction shifts toward higher values of $k$, since for each model the expansion after inflation (before setting to matter-dominated) is not exactly the same. Further, we observe that (1) as $\alpha$ decreases, more modes enter the IB, broadening the distribution, and (2) these modes grow faster, leading to a lower threshold $\delta_c^{\text{min}}(k)$, and consequently to a higher $\beta(k)$. {This last can be understood by inspecting the properties of the complementary error function. That is, if the threshold is sufficiently small so that $\delta_c^{\text{min}}(k)\lesssim\sigma_k$ (as it is the case for small $\alpha$ or long preheating phase \cite{Jedamzik:2010dq,Martin:2019nuw}), then $\beta(k)\simeq\erfc[\delta_c^{\text{min}}(k)/(\sqrt{2}\sigma_k)]\simeq\mathcal{O}(1)$ even for small amplification of perturbations. This is in contrast to other studies where the threshold is a fixed quantity of $\mathcal{O}(1)$, and the perturbations are required to be hugely amplified to produce a significant abundance of PBH. We emphasize again that this is not the standard implementation of the PS formalism. As explained above, we follow the approach of \cite{Martin:2019nuw}, whose differences with the conventional PS prescription are discussed in more detail in Appendix~\ref{app:appendixA}.
}

Overall, these effects result in a broader mass fraction that grows for smaller values of $\alpha$. If one increases the duration of preheating, we have checked that the value of $\alpha$ becomes irrelevant since almost the majority of the perturbations have enough time to collapse, resulting in an overproduction of PBH. This occurs for a duration of preheating of approximately 12-13 e-folds since at this stage the IB broadens to the point of including the modes belonging to the non-linear regime of the power spectrum {and also the threshold decreases drastically}. In that case, the mass fraction $\beta(k)$ becomes flat and fixed at 1 for the {altered-}PS formalism, independently of $\alpha$. This is what we refer to as an overproduction of PBH.

{In the literature, one can find estimations for the threshold coming from the Compaction Function approach \cite{Shibata:1999zs, Harada:2015yda,Harada:2023ffo,Musco:2018rwt,Escriva:2019phb,Harada:2024trx,Young:2024jsu}. However, as we explain in Appendix~\ref{app:AppendixB}, these thresholds are obtained numerically and, to the best of our knowledge, the studies focus almost exclusively on radiation-dominated backgrounds. For this reason, we do not take into account such an approach.}

\paragraph{Khlopov-Polnarev}

The next model we consider is the Khlopov-Polnarev (KP) formalism, independent of any collapse threshold and originally formulated for a matter-dominated universe, where the effects of pressure in halting the collapse are negligible. Particularly, in this regime, Carr's criterion for the threshold \cite{Carr:1975}, $\delta_c \sim w \sim 0$, implies that any slightly overdense region could collapse. However, this approach overestimates the abundance of PBH as it neglects nonspherical effects, which can prevent collapse. As the region containing the perturbation shrinks, any initial deviations from spherical symmetry can grow significantly, thereby avoiding collapse. To address this issue, Khlopov and Polnarev proposed in the 1980s \cite{Khlopov:1980,Khlopov:1981,Khlopov:1982} that the probability of PBH formation depends on the fraction of regions that are sufficiently spherical to undergo collapse. For instance, one can consider the effect that the anisotropy of the perturbations has on the collapse into a PBH. This analysis employs the Zel’dovich approximation to describe the nonlinear evolution of density perturbations, Thorne’s hoop conjecture, and the probability distribution for nonspherical perturbations derived by Doroshkevich  \cite{Doroshkevich:1970}. A semi-analytical refinement of this analysis, given in \cite{Harada:2016mhb}, leads to the following formula for small perturbations:
\begin{equation} \label{eq:KP-formula}
    \beta(k)\simeq0.056\sigma_k^{5},
\end{equation}
based solely on the anisotropy criterion. This expression is valid for $\sigma_k<0.01$ and thus applies to our case of study, since although the density perturbations amplify, the mass fraction is evaluated at the moment the perturbation crosses the IB and thus starts to amplify from a small value. {In fact, this is one of the main differences with respect to the altered-PS formalism. In this case, $\beta(k)\propto\sigma_k^5$ and only hugely initially amplified perturbations will result in high mass fractions. This explains the large difference (in order of magnitude) between {altered-}PS and KP in Fig.~\ref{fig:beta}.} In physical terms, the effect of the anisotropy can be understood as follows. In an almost spherical collapse, gravity pulls matter radially inward toward the center, but in an anisotropic collapse, matter collapses faster in some directions than others. If these differences are significant, shear stresses can disrupt the formation of a PBH \cite{Barrow:1978}. However, a moderate anisotropy can allow collapse. For instance, if a perturbation is slightly elongated or deformed but still retains a strong central gravitational potential, it can collapse into a PBH.

In Fig.~\ref{fig:beta}, the mass fraction computed using the KP formalism (continuous curves) is shown for both T- and E-models and various values of $\alpha$. The three criteria defined in Sec.~\ref{sec:pbh-formation} are used to determine the range of modes that can collapse, with calculations performed at 10 e-folds after the end of inflation. The results show a tendency similar to the {altered-}PS formalism: as $\alpha$ decreases, the mass fraction increases and shifts to higher values of $k$. However, a notable distinction is observed: in contrast to the {altered-}PS formalism, the KP mass fraction decreases with decreasing $\alpha$, for the same value of $k$. However, the modes affected cover a higher range due to the broadened IB and the smaller threshold from the fast amplification. Furthermore, in this case, the effect of increasing the duration of preheating does not necessarily translate into an overproduction of PBH, as the mass fraction $\beta(k)$ broadens but remains small, in general. However, as discussed in the {altered-}PS case, the modes belonging to the non-linear regime of the power spectrum enter the IB, and thus, it is still not very clear if one can trust the KP estimation in this scenario.

One can observe differences in the mass fractions with respect to \cite{del-Corral:2023}, particularly in the case of the KP formalism. This is because we are following a different approach in terms of the collapse time. In \cite{del-Corral:2023}, the collapse time is considered for all modes to be at the end of preheating when the density contrast has grown due to the parametric instabilities in Starobinsky inflation, which translates into a higher mass fraction. Some studies adopt this approach \cite{Niemeyer:1997mt,Niemeyer:1999,Green:1999xm,Gow:2020bzo}. In contrast, we consider a distinct collapse time for each mode, evaluating the density contrast at the moment of entering the IB, as some other works do \cite{Martin:2019nuw,Jedamzik:2010dq,Martin:2019nuw}. Despite these methodological differences, we find it valuable to explore both alternatives, given the limited research on PBH formation during early matter-dominated eras such as preheating. Further investigation in this area is essential to improve our understanding of the collapse of perturbations in these scenarios.

Further non-spherical effects can be taken into account to study the collapse under the KP formalism, such as the effect of the inhomogeneity \cite{Khlopov:1981,Kokubu:2018fxy,Harada:2016mhb} or the angular momentum of the black hole \cite{Harada:2017}. However, we restrict to the anisotropy criterion, as it is the dominant one. If all the criteria are considered, the total mass fraction is estimated as the product of the mass fractions associated with each effect, that is:
\begin{equation}
    \beta_\text{tot}=\beta_\text{aniso}\times\beta_\text{inhom}\times\beta_\text{spin}.
\end{equation}


\subsection{PBH mass}\label{sec:PBH-mass}

To compute the mass of the PBH forming, we estimate it roughly as the horizon mass $M_H$ at the moment these modes enter the IB \cite{Martin:2019nuw}, that is:
\begin{equation}\label{eq:PBH-mass}
    M_{\text{PBH}}=\gamma M_{H},
\end{equation}
where the parameter $\gamma$ specifies the fraction of the horizon mass that ends up in the PBH. It represents the uncertainties of the collapse in matter-dominated scenarios, and for simplicity, we take it to be $\gamma=1$. The horizon mass can be easily estimate as follows
\begin{equation}\label{eq:horizon-mass}
    M_{\text{H}}= \frac{4\pi}{3} \rho H^{-3}\simeq \frac{4\pi}{3} \rho_{end} {a_{end}^3}\left(\frac{1}{aH}\right)^3\, .
\end{equation}
This estimation for $M_{\text{PBH}}$ also differs from the one in \cite{del-Corral:2023}, where the critical scaling model is used instead \cite{Niemeyer:1999}. Numerical general relativistic simulations \cite{Maison:1995cc,Niemeyer:1999ak,Niemeyer:1997mt,Neilsen:1998qc,Musco:2012au,Snajdr:2005pr} have demonstrated that the scaling of black hole mass with the proximity to the formation threshold, known to occur in near-critical gravitational collapse, also applies to PBH formation. Under these assumptions, the PBH mass is given by:
\begin{equation}\label{eq:scaling}
    M_{\text{PBH}}(k)=M_{H}\kappa(\delta_{\bm k}-\delta_c)^{\gamma},
\end{equation}
where $\kappa$ and $\gamma$ are constants. However, in the present study, the differences $(\delta_{\bm k}-\delta_c(k))\gg1$ are due to the rapid growth of perturbations (which lowers the value of the thresholds). In this sense, perturbations are in the super-critical regime, where the critical scaling does not apply. For this reason, we compute the PBH mass following \eqref{eq:PBH-mass}

\begin{figure}[htbp]
    \centering
    \subfigure[]{%
    \includegraphics[width=0.95\linewidth]{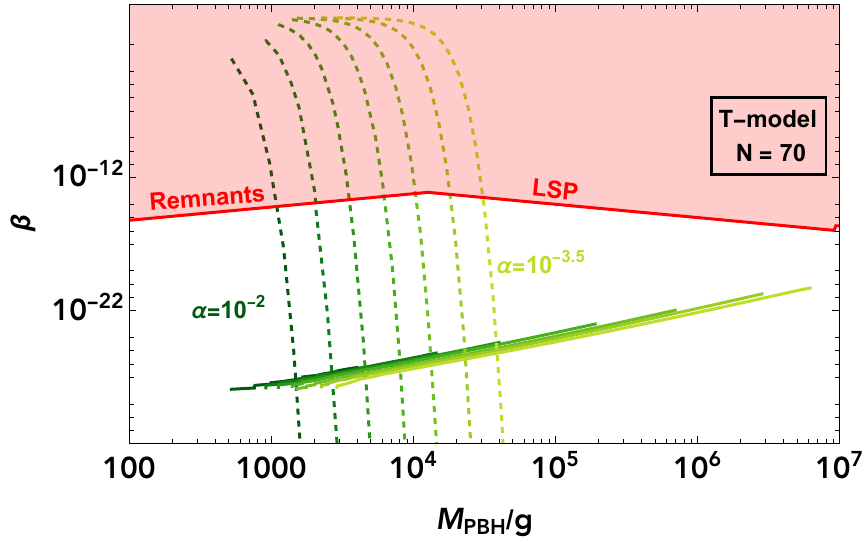}
        \label{fig:T-model-M-75}
    }
    \hfill
    \subfigure[]{%
        \includegraphics[width=0.95\linewidth]{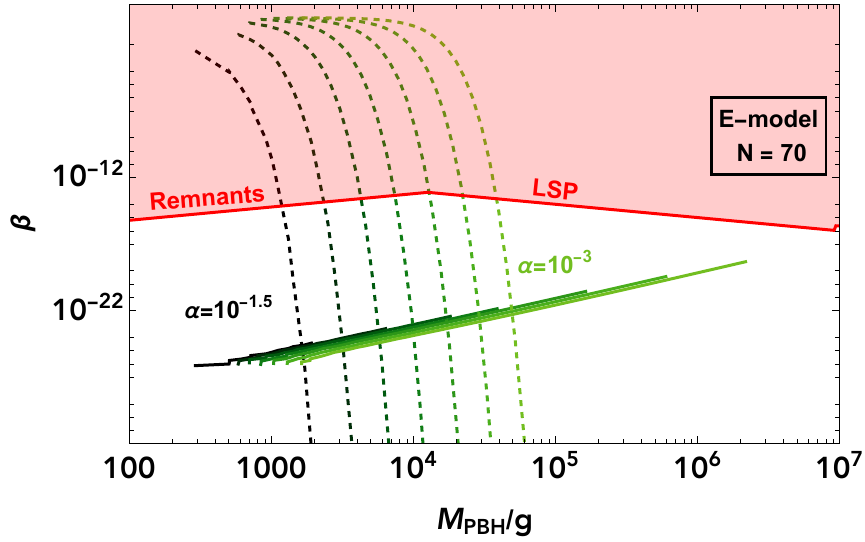}
        \label{fig:E-model-M-75}
    }
    \caption{Mass fraction for \textbf{(a)} T-model and \textbf{(b)} E-model using {altered-}PS (dashed) and KP (continuous) formalisms as a function of the mass of the PBH formed, eqn.~\eqref{eq:scaling}. Evaluations are made at 10 e-folds after the end of inflation. The red-shaded region represents the values of $\beta(k)$ constrained by the remnants and LSP constraints. The color code for $\alpha$ is the same as in Fig.~\ref{fig:beta}.}
    \label{fig:betaM}
\end{figure}

Figure~\ref{fig:betaM} displays the mass fraction as a function of the PBH mass for different values of the parameter $\alpha$ and for both the T- and E-models, using the {altered-}PS and KP formalisms. The effect of the parameter $\alpha$ is also noteworthy here; a decrease in $\alpha$ leads to higher PBH masses. Furthermore, the range of masses of the PBH formed is affected only by the constraints of Planck remnants and LSP, which, as stated in the introduction, are highly theoretical. Particularly, the {altered-}PS formalism is highly disfavored in this scenario if one assumes that Hawking radiation produces stable Planck remnants as the end products of evaporation. In contrast, the KP estimation is consistent with all the current constraints. Further, these significant differences between the {altered-}PS and KP formalisms emphasize the importance of accounting for nonspherical effects during the collapse of perturbations. Neglecting such effects can lead to an overestimation of the mass fraction or a rule-out of the model. Due to the lack of a definitive method, this analysis highlights the further need for accurate numerical simulations of PBH formation during preheating.


\section{Conclusions}\label{sec:conclusions}

In this work, we have explored the formation of PBH during the preheating phase of the early universe in the context of $\alpha$-attractor models of inflation \cite{Kallosh:2013yoa,Kallosh:2015,Iacconi:2023mnw,Carrasco:2015uma}, focusing on their mass fraction and mass spectrum under the frameworks of the {altered} Press-Schechter (PS) \cite{Press:1973,Harada:2013} and Khlopov-Polnarev (KP) \cite{Khlopov:1980,Khlopov:1981,Khlopov:1982} formalisms.  We found effects of self-resonance during the preheating that are particularly strong for $\alpha\ll1$ \cite{self-resonance} and induce instabilities that could lead to the collapse of overdensities and form PBH.  Using Floquet theory, we define a band of modes affected by the self-resonance effects, which we call the instability band (IB). By incorporating the dynamics of the density perturbations within the IB, we have defined three essential criteria that the density perturbations must fulfill to collapse into a PBH. Namely, (i) the co-moving wavenumber of the fluctuations must lie within the IB \eqref{eq:IB-criterion}, (ii) the wavelength of fluctuations must be sub-horizon and above the effective Jeans' length \eqref{eq:jeans-length}, and {(iii) the density contrast criterion \eqref{eq:time-of-collapse}, which states that an overdensity $\delta_{\bm k}$ associated with a scale has to be sufficient enough for the perturbations to collapse before the end of preheating, which is based on the collapse model of the massive scalar field in Eintein de Sitter universe \cite{Goncalves:2000nz,Martin:2019nuw}}. Our approach not only refines previous analyses \cite{Martin:2019nuw,Jedamzik:2010dq,Martin:2019nuw} but also crucially extends our previous results in the context of Starobinsky inflation \cite{del-Corral:2023} to all the Starobinsky-like scenarios like $\alpha$-attractor models. The major highlight of our study is that we considered the impact of small-scale instabilities and non-spherical effects in the formation of PBH by incorporating the KP formalism for the first time into generalized Starobinsky-like models. We quantitatively compared the PBH mass fraction $\beta(k)$ estimates from the {altered-}PS formalism, which is widely followed, against the KP formalism.

Using the {altered-}PS formalism \eqref{eq:modified-PS}, we computed the mass fraction $\beta(k)$ of PBH, taking into account Gaussian statistics for perturbations and a $k$-dependent variance derived from the power spectrum. Our results are shown in Fig.~\ref{fig:beta} and indicate that, as the parameter $\alpha$ decreases, the mass fraction shifts to higher values of $k$ and broadens. However, the {altered-}PS formalism overestimates PBH abundance in a matter-dominated universe, as it neglects nonspherical effects (although it considers a threshold value for the perturbations). On the other hand, the KP formalism \eqref{eq:KP-formula}, which accounts for the anisotropy criterion and is independent of any threshold, offers a more realistic estimate of PBH formation during preheating. Our analysis shows in Fig.~\ref{fig:beta} that the KP estimation is, in general, lower compared to the {altered-}PS case. {This can also be understood in terms of the threshold $\delta_c^{\text{min}}$ defined in eqn.~\eqref{eq:formal-threshold}. As explained previously, in some cases, it can be very low. In the {altered-}PS formalism, perturbations for which $\delta_{\bm k}\sim\delta_c$ imply $\beta(k)\sim\mathcal{O}(1)$, whereas for the KP formalism, the mass fraction is threshold-independent and just proportional to a power of the density contrast, see Eqn.~\eqref{eq:KP-formula}.} Also, the tendency with $\alpha$ differs from the {altered-}PS formalism. Smaller values of $\alpha$ lead to a smaller abundance of PBH for the same value of $k$, although it affects a broad range of modes. This analysis is slightly different from \cite{del-Corral:2023}, where the formation of PBH by the overdensities of all $k$ is considered to occur at the end of the preheating phase. In this paper, we considered different times of PBH formation depending on the minimum time for each fluctuation to collapse and the corresponding threshold of density contrast for each $k$. To be precise, the former approach considers an instant of time for PBH formation, i.e., the end of preheating, whereas our investigation in this paper explores different times of PBH formation for each scale before the end of preheating.  We find both approaches interesting and offer new insights into the phenomenon of PBH formation. 

Finally, we compared our results with observational constraints \cite{Carr:2020,Carr:2021bzv,Carr:1994ar,Zeldovich,Acharya:2020jbv,Carr:2009jm,Josan:2009qn,Lemoine:2000sq,Gondolo:2020uqv}, such as those from PBH remnants, LSP {and DM} production, CMB effects, BBN, and $\gamma$-ray backgrounds (see Fig.~\ref{fig:constraints} or red-shaded region in Fig.~\ref{fig:betaM}). For this particular scenario, the range of masses is only affected by the {Planck scale} remnants, LSP, {and DM} constraints, which are highly theoretical, {and the last two of them actually depend on the mass of the particle emitted by the PBH}. Still, our analysis demonstrates that the interplay between the non-spherical and self-resonance effects during preheating plays a crucial role in determining the formation of PBH.


\acknowledgments
Daniel del-Corral is grateful for the support of the grant UI/BD/151491/2021 from the Portuguese Agency Funda\c{c}\~ao para a Ci\^encia e a Tecnologia and the grant UMO-2021/42/E/ST9/00260 from the National Science Centre of Poland. This research was also funded by Funda\c{c}\~ao para a Ci\^encia e a Tecnologia grant number UIDB/MAT/00212/2020 and COST action 23130.  The work of P.G. was partially supported
by NSF grant PHY-2412829. P.G. thanks Prof. Teruaki Suyama for his kind hospitality
at the Institute of Science Tokyo. KSK would like to thank The Royal Society for the support in the name of the Newton International Fellowship. KSK would like to thank Prof. David Wands for useful discussions.

\appendix 

\section{Standard vs altered Press-Schechter formalism}\label{app:appendixA}

In this appendix, we clarify how the standard PS formalism is generically applied to a matter-dominated phase, such as preheating, and show the differences with the approach of \cite{Martin:2019nuw}, which we call altered-PS. 

\begin{figure}[t]
    \centering
    \subfigure[]{%
    \includegraphics[width=0.95\linewidth]{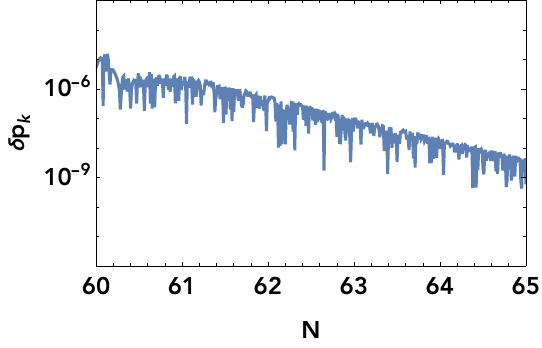}
        \label{fig:pressure1}
    }
    \hfill
    \subfigure[]{%
        \includegraphics[width=0.95\linewidth]{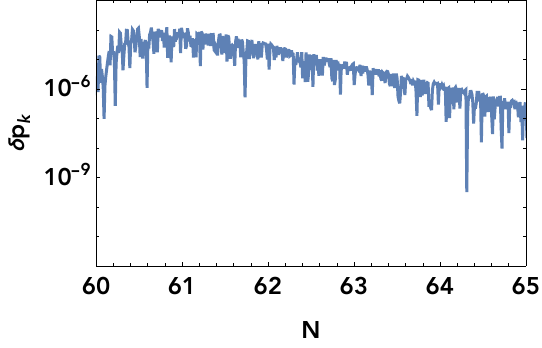}
        \label{fig:pressure2}
    }
    \caption{Pressure fluctuations for an E-model with \textbf{a)} $\alpha=10^{-2}$ and \textbf{b)} $\alpha=10^{-3}$. The wavenumber $k$ corresponds to the one that maximizes the amplification of the curvature perturbations.}
    \label{fig:pressure}
\end{figure}

The preheating phase is considered to be effectively (or averaged) matter-dominated because the equation-of-state parameter $w$, although oscillating, averages to zero. But, to be more precise, we show that this is the case even at the perturbative level. Fig.~\ref{fig:pressure}, shows the evolution of the pressure fluctuations for an E-model with $\alpha=10^{-2}$ and $\alpha=10^{-3}$ for the wavenumbers of maximum amplification of the curvature perturbations. One can observe that these fluctuations start with a relatively small amplitude by the end of inflation, but as preheating evolves, the universe enters into the effective matter-dominated {stage}. As the parameter $\alpha$ decreases, we can observe that pressure fluctuations also {show} some kind of amplification, but it is not comparable to the case of the curvature fluctuations. 
Thus, the preheating phase in those cases behaves as matter-dominated even at the perturbative level. In this scenario, one can employ the top-hat spherical collapse model {where} the threshold for {for an overdensity to} collapse {is known to be} $\delta_c^{\text{lin}}\simeq1.686$ \cite{Press:1973,Peebles1980}. This is computed as follows. Inside the overdense region, we assume a closed FLRW metric
\begin{equation}
\dd s^{2} = -\dd t^{2} + a_{p}^{2}(t)\left[ \frac{\dd r^{2}}{1 - \kappa r^{2}} + r^{2} \dd\Omega^{2} \right],
\end{equation}
filled with dust and with a positive curvature $\kappa > 0$. Here, $\dd\Omega^2$ is the metric on the 2-sphere. The Hubble rate inside this region is given by
\begin{equation}\label{eq:Friedmann-perturbation}
H_p^2= \left( \frac{\dot a_{p}}{a_{p}} \right)^{2}
= \frac{8\pi G}{3}\,\rho_{p} - \frac{\kappa}{a_{p}^{2}},
\end{equation}
where the suffix ``$_{p}$'' refers to quantities of the interior of the overdense (perturbed) region. Outside of it, we consider an Einstein de Sitter (EdS) universe, which is described by a flat (matter-dominated) FLRW metric, for which
\begin{equation}\label{eq:Friedmann-background}
H_b^2=\left( \frac{\dot a_b}{a_b} \right)^{2}
= \frac{8\pi G}{3}\,\rho_{b}, \qquad a_b(t) \propto t^{2/3},
\end{equation}
where now the suffix ``$_b$'' refers to the background quantities. The solution of \eqref{eq:Friedmann-perturbation} can be obtained as the following parametric cycloid equation
\begin{align}
a_{p}(\theta) & = \frac{A}{2\kappa}\,\big(1 - \cos\theta\big), \label{eq:parametrization-scale-factor}\\ 
t(\theta) & = \frac{A}{2\kappa^{3/2}}\big(\theta-\sin\theta\big),\label{eq:parametrization-time}
\end{align}
where $\theta\in[0,2\pi]$ parametrizes the evolution of the overdense region. The point $\theta_0=0$, for which $a_p(\theta_0)=a_b(\theta_0)=0$ and $t(\theta_0)=0$, corresponds to the ``\textit{Big-Bang time}'', {close to the time scale of when} inflation begins. This is analogous to ignoring the radiation-dominated era in dark halo growth in standard cosmologies. From this point onwards, both regions evolve in the same way. In eqn.~\eqref{eq:parametrization-time} we could have added an initial time $t_0$ so that $t(\theta_0)=t_0$, but this will just introduce a time shift that complicates the formulae. In any case, the point $\theta=t=0$ is never physically reached to obtain $\delta_c^{\text{lin}}$, only the limit $\theta\ll1$. After inflation, the perturbed region enters the IB at $\theta_{\text{bc}}$, or $t_{\text{bc}}$ in cosmic time, where $a_p(\theta_{\text{bc}})=a_b(t_\text{bc})>0$. This is the moment when the perturbed region detaches from the Hubble flow and its growth slows. The value $\theta_{\text{ta}}=\pi$ is the turnaround of the overdensity (the starting point of the contraction). Since the turnaround must occur after the perturbation enters the IB, we know that $0<\theta_{\text{bc}}<\theta_{\text{ta}}$, but the precise value of $\theta_{\text{bc}}$ is also not needed to estimate $\delta_c^{\text{lin}}$. The cycloid solution \eqref{eq:parametrization-scale-factor} and \eqref{eq:parametrization-time} is just a parametrization of the evolution of the perturbed universe that we use to obtain the critical density contrast. Finally, the moment $\theta_{\text{coll}}=2\pi$ corresponds to the collapse time, for which $a_p=0$ again. The constant $A$ is defined as
\begin{equation}
    A = \frac{8\pi G}{3}\rho_{p}(t_{\text{bc}})a_{p}^3(t_{\text{bc}}),
\end{equation}
and we are considering that the energy density of the perturbed region evolves as {pressureless matter given by}
\begin{equation}\label{eq:density-param}
    \rho_p=\rho_{p}(t_{\text{bc}})\left(\frac{a_{p}(t_{\text{bc}})}{a_p}\right)^3.
\end{equation}

In relation to the positive curvature $\kappa$, this can be estimated by evaluating eqns.~\eqref{eq:Friedmann-perturbation} and \eqref{eq:Friedmann-background} at the band-crossing time $t_{\text{bc}}$, obtaining
\begin{equation}\label{eq:kappa}
    \kappa=H_b^2(t_{\text{bc}})\delta_{\text{nl}}(t_{\text{bc}})a_p^2(t_{\text{bc}}),
\end{equation}
where $\delta_{\text{nl}}$ is the non-linear density contrast, defined similarly as in Sec.~\ref{sec:pbh-formation}, $\delta_{\text{nl}}=\frac{\rho_p-\rho_b}{\rho_b}=\frac{\delta\rho}{\rho_b}$, with $\delta\rho$ being the density perturbation itself. We can now evaluate eqn.~\eqref{eq:parametrization-time} at $\theta_{\text{coll}}=2\pi$ and use eqn.~\eqref{eq:kappa} to obtain the collapse time in cosmic time
\begin{equation}\label{eq:time-of-collapse-2}
    t_{\text{coll}}=\frac{\pi}{H_b(t_{\text{bc}})\delta_{\text{nl}}^{3/2}(t_{\text{bc}})}.
\end{equation}
We can identify this equation with eqn.~\eqref{eq:time-of-collapse} and with eqn.~(B.7) of \cite{Martin:2019nuw}.

Let us compute the evolution of the non-linear density contrast $\delta_{\text{nl}}$, which we first rewrite as
\begin{equation}\label{evolution-overdensity}
    \delta_{\text{nl}}=\frac{\rho_p-\rho_p}{\rho_b}=\frac{\rho_p}{\rho_b}-1=\left(\frac{a_b}{a_p}\right)^3-1,
\end{equation}
where in the last step we have parametrized $\rho_b$ as in Eqn.~\eqref{eq:density-param}. Using Eqn.~\eqref{eq:parametrization-time}, we know that the background scale factor evolves during matter-domination as
\begin{equation}
\begin{split}
a_b&=a_b(t_{\text{bc}})\left(\frac{t}{t_{\text{bc}}}\right)^{2/3}\\&=a_b(t_{\text{bc}})\left(\frac{A}{2\kappa^{3/2}t_{\text{bc}}}\bigl(\theta-\sin\theta\bigr)\right)^{2/3}.
\end{split}
\end{equation}
Therefore, substituting this and the Eqn.~\eqref{eq:parametrization-scale-factor} into Eqn.~\eqref{evolution-overdensity}, we obtain, after some algebra, the following expression for the non-linear density contrast
\begin{equation}
    \delta_{\text{nl}}=\frac92\frac{\bigl(\theta-\sin\theta\bigr)^2}{\bigl(1-\cos\theta\bigr)^3}-1.
\end{equation}
However, in our computations, we employ the linear density contrast, which we know evolves as $\delta\sim a_b\sim t^{2/3}$. Thus, we can temporarily parametrize it as
\begin{equation}
    \delta_{\text{lin}}\simeq C\, \bigl(\theta - \sin\theta\bigr)^{2/3},
\end{equation}
where we have used again Eqn.~\eqref{eq:parametrization-time} and $C$ is a constant. By matching both density contrasts at early times (expanding around $\theta\sim0$), we found that the constant $C$ is given by
\begin{equation}
    C=\frac{3}{20}6^{2/3}.
\end{equation}
Then, the expression for the linear density contrast can be derived as
\begin{equation}
    \delta_{\text{lin}}=\frac{3}{20}\bigl[6(\theta - \sin\theta\bigr)]^{2/3},
\end{equation}
which when evaluated at the collapse time $\theta_{\text{coll}}=2\pi$ gives
\begin{equation}
    \delta_c^{\text{lin}}=\frac{3(12\pi)^{3/2}}{20}\simeq1.686.
\end{equation}
{Thus, i}n the standard PS formalism, Eqn.~\eqref{eq:PS-formalism}, one evaluates the quotient
\begin{equation}\label{eq:ratio}
\frac{\delta_c^{\text{lin}}}{\sqrt{2}\,\sigma_k}\simeq \frac{\delta_c^{\text{lin}}}{\sqrt{2\mathcal{P}_{\delta}(k)}}.
\end{equation}
Thus, the closer the fluctuations are to the threshold, the smaller this ratio becomes. Then, because of the behavior of the $\erfc$ function, this leads to a larger mass fraction. On the contrary, a large ratio implies a small mass fraction. Using $\delta_c=\delta_c^{\text{lin}}\simeq1.686$ in the standard PS formula \eqref{eq:PS-formalism} makes the ratio \eqref{eq:ratio} extremely large, suppressing $\beta(k)$ to essentially zero. In that case, the standard-PS formalism predicts no PBH formation.

To circumvent this, the authors of \cite{Martin:2019nuw} alter the standard PS formalism as follows. Using the top-hat spherical collapse model described above, the authors derive the collapse time of a perturbation, given by ~\eqref{eq:time-of-collapse}. This corresponds to the time required for the linear density contrast, $\delta_{\text{lin}}$, to grow to order unity, at which point the collapse is assumed to occur. Depending on the preheating duration, some perturbations may have sufficient time to collapse. This allows one to determine, for each wavenumber $k$, a lower bound on the initial density contrast $\delta_{\bm k}$ for the collapse to occur. This lower bound is what they (and we) refer to as the threshold $\delta_c^{\text{min}}(k)$, given by eq.~\eqref{eq:formal-threshold}. This is essentially what we refer to as the altered-PS formalism, defined now in Eqn.~\eqref{eq:modified-PS}. 

By inspecting the equation \eqref{eq:formal-threshold} of the threshold $\delta_c^{\text{min}}$, depending on the length of preheating, $N_{\text{rh}}$, and the band-crossing time of each mode, $N_{\text{bc}}(k)$, the threshold can be as small as $\mathcal{O}(10^{-5})$, which is comparable to the typical amplitude of the fluctuations when they enter the IB. This in turn means that the ratio \eqref{eq:ratio} (with now $\delta_c^{\text{lin}}$ substituted by $\delta_c^{\text{min}}$) is now small and therefore the mass fraction $\beta(k)$ is large. This explains why the altered-PS formalism yields comparatively large estimates for $\beta(k)$, in huge contrast to the KP formalism, in which $\beta(k)$ scales as a power of the fluctuation amplitude, see Eqn.~\eqref{eq:KP-formula}. See also Figs.~\ref{fig:beta} and \ref{fig:betaM} for comparisons between these formalisms.

As shown, the standard PS approach does not predict PBH formation in this context. However, adopting the altered-PS procedure of \cite{Martin:2019nuw} reveals a clear trend: PBH production increases as the instabilities strengthen (i.e., as $\alpha$ decreases). Since highlighting this behavior is a primary aim of the present work, we regard the use of this alternative implementation of the PS formalism as both illustrative and informative.


\section{Compaction function formalism in matter-dominated scenarios}\label{app:AppendixB}

The compaction function was originally introduced by Shibata and Sasaki in the context of numerical-relativity studies of PBH formation \cite{Shibata:1999zs}. In their formulation, the compaction was defined on a constant-mean-curvature (CMC) slicing and constructed as a measure of the mass excess within a given radius, normalized by the areal radius. Physically, it quantifies the mass concentration within a region relative to a homogeneous FLRW background, thereby providing a direct probe of the gravitational strength of the perturbation. More precisely, the original Shibata–Sasaki compaction function is defined as an integral of the density perturbation on a CMC hypersurface. This definition is slicing-dependent, but empirical evidence \cite{Shibata:1999zs} shows that it provides a robust criterion for PBH formation. This result was obtained through numerical-relativity simulations of spherically symmetric perturbations, highlighting that the threshold is intrinsically dynamical and cannot be derived purely analytically. Subsequent works reformulated the compaction function in a more geometrical and physically transparent way \cite{Harada:2015yda,Harada:2023ffo,Musco:2018rwt,Escriva:2019phb}. In particular, it was recognized that a more natural definition is given by
\begin{equation}
C(r) = \frac{\delta M(r)}{R(r)},
\end{equation}
where $\delta M$ is the Misner-Sharp mass excess and $R$ is the areal radius. This definition is covariant under spatial coordinate transformations and provides a quasi-local measure of the overdensity. Importantly, while it remains slicing-dependent, it is much more directly connected to the dynamics of gravitational collapse, since the Misner-Sharp mass is a geometrically well-defined quantity in spherical symmetry. 

A key point clarified in later works \cite{Harada:2024trx} is that different definitions of the compaction function (e.g.~the original Shibata–Sasaki version and the Misner–Sharp-based one in comoving slicing) are not strictly identical. However, they are related in the long-wavelength limit and for simple equations of state. In particular, for a perfect fluid with $p = w \rho$, the two definitions differ by a constant factor depending on $w$. This highlights that part of the apparent ambiguity in the literature is due to gauge choices and to the specific variables used to characterize the perturbation. Therefore, the different thresholds quoted in the literature should not be interpreted as reflecting different physics, but rather as arising from different choices of variables, parametrizations, and equations of state. For instance, the dependence on the equation of state arises from the competition between gravitational attraction and pressure gradients: softer equations of state (smaller $w$) reduce pressure support and therefore lower the collapse threshold. In fact, an important point emphasized in the literature is that different amplitude measures (such as the averaged density contrast $\tilde{\delta}$, the compaction $C_{\rm max}$, or the central curvature perturbation) can exhibit different sensitivities to the actual shape of the curvature (or density) profile $K(r)$ and therefore to the surrounding environment \cite{Shibata:1999zs,Harada:2015yda,Young:2024jsu}. Physically, this curvature profile describes how spatial slices deviate from flatness due to the presence of an overdensity. A positive curvature region corresponds to a locally closed geometry, indicating that the region contains more mass than the background and is therefore prone to gravitational collapse. The compaction function can be directly related to the curvature profile in this regime, showing that both quantities encode the same physical information in different variables \cite{Harada:2024trx}. In particular, profiles with sharper transitions between the overdense region and the surrounding background exhibit stronger pressure gradients, which oppose collapse and increase the threshold. Conversely, smoother profiles reduce pressure gradients and lead to lower threshold values \cite{Harada:2015yda}. This dependence has been confirmed both analytically (in simplified models) and numerically for a wide class of profiles.

The compaction function provides a physically well-motivated criterion for PBH formation, as it is directly related to the integrated mass excess responsible for collapse and is less sensitive to large-scale environmental effects than local quantities. For this reason, it is particularly well-suited for estimating PBH abundances. However, the critical threshold within the compaction function framework is calibrated through numerical-relativity simulations of spherically symmetric perturbations. To date, such calibrations have been performed almost exclusively for radiation dominated backgrounds, where extensive numerical studies exist and have established robust threshold values \cite{Shibata:1999zs,Harada:2015yda,Harada:2024trx,Young:2024jsu,Musco:2018rwt,Escriva:2019phb,Young:2024jsu,Harada:2023ffo}. To the best of our knowledge, there are currently no dedicated numerical-relativity studies that determine a compaction function threshold in a matter-dominated universe. Therefore, while thresholds obtained using the spherical-collapse approach (Appendix \ref{app:appendixA}) and the compaction function framework can be meaningfully compared in radiation domination, the absence of an established compaction-based threshold in matter-dominated scenarios prevents a direct and consistent comparison in that case. In particular, the value of $\delta_c(k)$ used in our analysis is defined within the spherical-collapse framework and is not directly mapped to a compaction threshold.

Finally, we emphasize that although different formalisms (compaction function, averaged density contrast, curvature perturbation) provide different parametrizations of the collapse condition, they ultimately describe the same physical process. The main limitation in the matter-dominated case is, in principle, not conceptual, but rather the lack of numerical calibration of the compaction-based criterion in that regime. We refer this to a future work.


\bibliographystyle{apsrev4-2}
\bibliography{BIBLIO.bib}

\end{document}